\documentclass[preprint,prd,nofootinbib,tightenlines,groupedaddress,amsmath,amssymb]{revtex4}

\usepackage{bm}
\usepackage{color}
\usepackage{graphicx}
\usepackage{multirow}

\begin{document}
\baselineskip=15pt \parskip=3pt

\vspace*{3em}

\title{Fermion EDMs with Minimal Flavor Violation}

\author{Xiao-Gang He,$^{1,2,3}$ Chao-Jung Lee,$^{3}$ Siao-Fong Li,$^{3}$ and Jusak Tandean$^{3}$}
\affiliation{${}^{1}$INPAC, SKLPPC, and Department of Physics,
Shanghai Jiao Tong University, Shanghai 200240, China \vspace{3pt} \\
${}^{2}$Physics Division, National Center for
Theoretical Sciences, Department of Physics, National Tsing Hua
University, Hsinchu 300, Taiwan \vspace{3pt} \\
${}^{3}$CTS, CASTS, and
Department of Physics, National Taiwan University, Taipei 106, Taiwan \vspace{3ex}}


\begin{abstract}
We study the electric dipole moments (EDMs) of fermions in the standard model
supplemented with right-handed neutrinos and its extension including the neutrino
seesaw mechanism under the framework of minimal flavor violation (MFV).
In the quark sector, we find that the current experimental bound on the neutron EDM
does not yield a~significant restriction on the scale of MFV.
In addition, we consider how MFV may affect the contribution of the strong
theta-term to the neutron EDM.
For the leptons, the existing EDM data also do not lead to strict limits if
neutrinos are Dirac particles.
On the other hand, if neutrinos are Majorana in nature, we find that the constraints
become substantially stronger.
Moreover, the results of the latest search for the electron EDM by the ACME
Collaboration are sensitive to the MFV scale of order a~few hundred GeV or higher.
We also look at constraints from $CP$-violating electron-nucleon interactions that
have been probed in atomic and molecular EDM searches.
\end{abstract}

\maketitle

\section{Introduction\label{intro}}

Searches for electric dipole moments (EDMs) are a powerful means of probing new sources
of the violation of charge parity $(CP)$ and time reversal $(T)$ symmetries beyond
the standard model~(SM) of particle physics~\cite{He:1989xj,suzuki,Ginges:2003qt,Engel:2013lsa}.
Recently the ACME experiment~\cite{acme}, which utilized the polar molecule thorium monoxide
to look for the EDM of the electron, $d_e$, has produced a~new result of
\,$d_e=(-2.1\pm3.7_{\rm stat} \pm 2.5_{\rm syst})\times 10^{-29\;}e$\,cm,\, which corresponds
to an upper limit of \,$|d_e|< 8.7\times 10^{-29}\,e$\,cm\, at 90\% confidence level (CL).
This is more stringent than the previous best bound by about an order of magnitude,
but still way above the SM expectation for~$d_e$, which is at the level
of \,$10^{-44\;}e$\,cm~\cite{Pospelov:2013sca}.
Hence there is abundant room between the current limit and SM value of $d_e$
where potential new physics may be observed in future measurements.
In the quark sector, the EDM of the neutron, $d_n$, plays an~analogous role in the quest of
new physics.
At present its experimental limit is \,$|d_n|<2.9\times10^{-26\;}e$\,cm\, at 90\% CL~\cite{pdg},
while the SM predicts it to be in the range of \,$10^{-32}$-$10^{-31\;}e$\,cm~\cite{hesm}.

Extra ingredients beyond the SM can increase the electron and neutron EDMs tremendously
with respect to their SM predictions, even up to their existing measured bounds.
Such substantial enlargement may have various causes which could greatly differ from model
to model.
It is, therefore, of interest to analyze fermion EDMs arising from possible nonstandard
origins under a~framework that allows one to deal with some general features of the physics
without getting into model specifics.
This turns out to be feasible under the context of the so-called minimal flavor violation
(MFV) which presupposes that the sources of all flavor-changing neutral currents (FCNC) and
$CP$ violation reside in renormalizable Yukawa couplings defined at tree
level~\cite{mfv1,D'Ambrosio:2002ex,Cirigliano:2005ck}.
Thus the MFV framework offers a systematic way to explore SM-related new interactions
which do not conserve flavor and $CP$ symmetries.

In an earlier paper~\cite{He:2014fva}, motivated by the recent ACME data, we have adopted
the MFV hypothesis in order to examine $d_e$ in the SM slightly expanded with
the inclusion of three right-handed neutrinos and in its extension incorporating the seesaw
mechanism for light neutrino mass generation.
In the present work, we would like to provide a~more extensive treatment of our previous study,
covering the EDMs of the other charged leptons as well.
For $d_e$ particularly, we demonstrate in greater detail how various factors may affect it
within the MFV context, taking into account extra empirical information on neutrino masses.
Moreover, we address the possibility that $d_e$ is correlated with the effective Majorana mass
that is testable in ongoing and upcoming searches for neutrinoless double-beta decay.
We will also perform an MFV analysis on the quark EDMs and estimate the resulting neutron~EDM.
In addition, we consider the MFV effect on the contribution of the theta term in QCD to
the neutron EDM.

The structure of the paper is as follows.
In Section \ref{mfv}, we describe the MFV framework and its aspects which are relevant to our
evaluation of fermion EDMs as probes for the scale of MFV.
In Section$\;$\ref{fedm} we derive the expressions for quark and lepton EDMs from several
effective operators satisfying the MFV principle.
Section$\;$\ref{numerics} contains our numerical analysis.
After determining the neutron EDM from the quark contributions and inferring the constraint
on the MFV scale from the neutron data, we examine how
the contribution of the QCD theta-term is altered in the presence of MFV.
In the lepton sector, we devote much of our attention to the electron EDM in light of
the ACME data and briefly address its muon and tau counterparts.
The acquired constraints on the MFV scale depend considerably on whether the light neutrinos
are Dirac or Majorana in nature, the EDMs in the former case being much smaller than the latter.
Subsequently, we look at $CP$-violating electron-nucleon interactions, which were also
investigated by ACME and other experiments looking for atomic or molecular EDMs.
Lastly, we discuss potential constraints from flavor-changing and other flavor-conserving
processes.
We make our conclusions in Section$\;$\ref{conclusion}.
An~appendix collects some useful lengthy formulas.

\section{Minimal flavor violation framework\label{mfv}}

In the SM supplemented with three right-handed neutrinos, the renormalizable Lagrangian for
the quark and lepton masses can be expressed as
\begin{eqnarray} \label{Lm}
{\cal L}_{\rm m}^{} &\,=\,& -\bar Q_{k,L\,}^{}(Y_u)_{kl\,}^{}U_{l,R\,}^{} \tilde H
- \bar Q_{k,L\,}^{}(Y_d)_{kl\,}^{}D_{l,R\,}^{} H
- \bar L_{k,L\,}^{}(Y_\nu)_{kl\,}^{}\nu_{l,R\,}^{}\tilde H
- \bar L_{k,L\,}^{}(Y_e)_{kl\,}^{}E_{l,R\,}^{} H
\nonumber \\ && \!-~
\mbox{$\frac{1}{2}$}\, \overline{\nu^{\rm c}}_{\!\!\!k,R}^{}\,(M_\nu)_{kl\,}^{}\nu_{l,R}^{}
\;+\; {\rm H.c.} ~,
\label{majorana-neu}
\end{eqnarray}
where summation over \,$k,l=1,2,3$\, is implicit, $Q_{k,L}$ $(L_{k,L})$ represents left-handed
quark (lepton) doublets, $U_{k,R}$ and $D_{k,R\,}$ $\bigl(\nu_{k,R}^{}$ and $E_{k,R}\bigr)$
denote right-handed up- and down-type quarks (neutrinos and charged leptons), respectively,
$Y_{u,d,\nu,e}$ are matrices containing the Yukawa couplings,
$M_\nu$~is the Majorana mass matrix of the right-handed neutrinos, $H$ is the Higgs doublet,
and \,$\tilde H=i\tau_2^{}H^*$\, involving the second Pauli matrix $\tau_2^{}$.
The Higgs' vacuum expectation value \,$v\simeq246$\,GeV\, breaks the electroweak symmetry as
usual, which makes the weak gauge bosons and charged leptons massive and also
induces Dirac mass terms for the neutrinos.
The $M_\nu$ part in ${\cal L}_{\rm m}$ plays an essential role in the type-I seesaw
mechanism~\cite{seesaw1}.\footnote{An analogous situation occurs in the type-III
seesaw model~\cite{seesaw3}.}
If neutrinos are Dirac particles, however, the $M_\nu$ terms are absent.

For the quark sector, the MFV hypothesis~\cite{D'Ambrosio:2002ex} implies that the Lagrangian
in Eq.\,(\ref{Lm}) is formally invariant under the global group
\,${\rm U}(3)_Q\times{\rm U}(3)_U\times{\rm U}(3)_{D} =
G_q\times{\rm U}(1)_Q\times{\rm U}(1)_U\times{\rm U}(1)_D$,\,
where \,$G_q = {\rm SU}(3)_Q\times{\rm SU}(3)_U\times{\rm SU}(3)_{D}$.\,
This entails that the three generations of $Q_{k,L}$, $U_{k,R}$, and $D_{k,R}$ transform as
fundamental representations of the SU$(3)_{Q,U,D}$, respectively, namely
\begin{eqnarray}
Q_L^{} \,\to\, V_Q^{}Q_L^{}~, ~~~~~~~ U_R^{} \,\to\, V_U^{}U_R^{}~, ~~~~~~~
D_R^{} \,\to\, V_D^{}D_R^{}~, ~~~~~~~ V_{Q,U,D} \,\in\, \rm SU(3) ~.
\end{eqnarray}
Moreover, the Yukawa couplings are taken to be spurions which transform according to
\begin{eqnarray}
Y_u^{} \,\to\, V_Q^{}Y_u^{}V^\dagger_U ~, ~~~~~~~ Y_d^{} \,\to\, V_Q^{}Y_d^{}V^\dagger_D ~.
\end{eqnarray}

Consequently, to arrange nontrivial FCNC and $CP$-violating interactions satisfying the MFV
principle and involving no more than two quarks, one puts together
an~arbitrary number of the Yukawa coupling matrices \,$Y_u\sim(3,\bar 3,1)$\, and
\,$Y_d\sim(3,1,\bar 3)$\, as well as their Hermitian conjugates to set up the $G_q$
representations \,$\Delta_q\sim(1\oplus8,1,1)$, \,$\Delta_{u8}\sim(1,1\oplus8,1)$,
\,$\Delta_{d8}\sim(1,1,1\oplus8)$, \,$\Delta_u\sim(\bar 3,3,1)$,\,
and~\,$\Delta_d\sim(\bar 3,1,3)$,\, combines them with two quark fields to build
the $G_q$-invariant objects \,$\bar Q_L\gamma_\alpha^{}\Delta_q Q_L$,
\,$\bar U_R\gamma_\alpha^{}\Delta_{u8}U_R$, \,$\bar D_R\gamma_\alpha^{}\Delta_{d8}D_R$,
\,$\bar U_R(1,\sigma_{\alpha\beta})\Delta_u Q_L$,\, and
\,$\bar D_R(1,\sigma_{\alpha\beta})\Delta_d Q_L$,\, includes appropriate numbers of the Higgs and
gauge fields to arrive at singlets under the SM gauge group, and contracts all the Lorentz indices.
Since \,$\bar Q_L\gamma_\alpha^{}\Delta_q Q_L$, \,$\bar U_R\gamma_\alpha^{}\Delta_{u8}U_R$,\,
and \,$\bar D_R\gamma_\alpha^{}\Delta_{d8}D_R$\, in this case must be Hermitian,
$\Delta_{q,u8,d8}$ must be Hermitian as well.

The Lagrangian describing the EDM $d_f^{}$ of a fermion $f$ is
\,${\cal L}_{f\,\rm edm}=-(i d_f/2)\bar f \sigma^{\kappa\omega}\gamma_5 f F_{\kappa\omega}$,\,
where $F_{\kappa\omega}$ is the photon field strength tensor.
Accordingly, among the combinations listed in the preceding paragraph, only
\,$\bar U_R^{}\sigma_{\alpha\beta}\Delta_u Q_L^{}$\, and
\,$\bar D_R^{}\sigma_{\alpha\beta}\Delta_d^{}Q_L^{}$\, pertain to our examination of quark EDMs.
For~$\Delta_{u,d}$, one can take \,$\Delta_u^{}=Y_u^\dagger\Delta^{}$\, and
\,$\Delta_d^{}=Y_d^\dagger\Delta^{}$,\, where $\Delta$ is built up of terms in
powers of \,${\sf A}=Y_u^{}Y_u^\dagger$\, and~\,${\sf B}=Y_d^{}Y^\dagger_d$,\,
which transform as $(1\oplus8,1,1)$ under $G_q$.

Formally $\Delta$ comprises an infinite number of terms, namely
\,$\Delta=\sum\xi_{jkl\cdots\,}^{}{\sf A}^j{\sf B}^k{\sf A}^l\cdots$\, with coefficients
$\xi_{jkl\cdots}^{}$ expected to be at most of~${\cal O}$(1).
The MFV hypothesis requires that $\xi_{jkl...}^{}$ be real because complex $\xi_{jkl...}^{}$
would introduce new $CP$-violation sources beyond that in the Yukawa couplings.
Using the Cayley-Hamilton identity
\begin{eqnarray} \label{ch}
X^3 \,\,=\,\, X^2\,{\rm Tr}X \,+\, \mbox{$\frac{1}{2}$}_{\,}X\bigl[{\rm Tr}X^2-({\rm Tr}X)^2\bigr]
\,+\, \openone{\rm Det}X
\end{eqnarray}
for an invertible 3$\times$3 matrix $X$, one can resum the infinite series into a finite
number of terms~\cite{Colangelo:2008qp,Mercolli:2009ns}
\begin{eqnarray}
\Delta &\,=\,& \xi^{}_1\openone + \xi^{}_{2\,}{\sf A}+\xi^{}_{3\,}{\sf B}
+ \xi^{}_{4\,}{\sf A}^2 +\xi^{}_{5\,}{\sf B}^2 + \xi^{}_{6\,}{\sf AB} + \xi^{}_{7\,}{\sf BA}
+ \xi^{}_{8\,}{\sf ABA} + \xi^{}_{9\,}{\sf BA}^2 + \xi^{}_{10\,}{\sf BAB}
\nonumber \\ && \! +~
\xi^{}_{11\,}{\sf AB}^2 + \xi^{}_{12\,}{\sf ABA}^2 + \xi^{}_{13\,}{\sf A}^2{\sf B}^2
+ \xi^{}_{14\,}{\sf B}^2{\sf A}^2 + \xi^{}_{15\,}{\sf B}^2{\sf AB}
+ \xi^{}_{16\,}{\sf AB}^2{\sf A}^2+\xi^{}_{17\,}{\sf B}^2{\sf A}^2{\sf B} ~, ~~~ \label{general}
\end{eqnarray}
where $\openone$ denotes a 3$\times$3 unit matrix.
One can then also utilize this to devise Hermitian combinations such as
\,$\Delta_q=\Delta+\Delta^\dagger$.\,

Even though one starts with all $\xi_{jkl\cdots\,}^{}$ being real, the resummation process will
render the coefficients $\xi_r^{}$ in Eq.\,(\ref{general}) generally complex due to imaginary
parts generated among the traces of the matrix products \,${\sf A}^j{\sf B}^k{\sf A}^l\cdots$\,
with \,$j+k+l+\cdots\ge6$\, upon the application of the Cayley-Hamilton identity.
In Appendix$\;$\ref{ABproducts} we show the detailed reduction of one of the lowest-order
products which give rise to the imaginary components of~$\xi_r^{}$.
We find that the imaginary contributions are always reducible to factors proportional to
\,${\rm Im\,Tr}\bigl({\sf A}^2{\sf BAB}^2\bigr)=(i/2)\,{\rm Det}[{\sf A,B}]$\,
which is a Jarlskog invariant and much smaller than one~\cite{Colangelo:2008qp}.

Taking advantage of the invariance under $G_q$, we will work in the basis where $Y_d$ is
diagonal,
\begin{eqnarray}
Y_d \,\,=\,\, \frac{\sqrt2}{v}\;{\rm diag}\bigl(m_d^{},m_s^{},m_b^{}\bigr) \,,
\end{eqnarray}
and the fields $U_{k,L}$, $U_{k,R}$, $D_{k,L}$, and $D_{k,R}$ belong to the mass eigenstates.
Hence we can write $Q_{k,L}$ and $Y_u$ in terms of the Cabibbo-Kobayashi-Maskawa (CKM)
quark mixing matrix $V_{\scriptscriptstyle\rm CKM}$ as
\begin{eqnarray} \label{Yu}
Q_{k,L}^{} \,\,=\,\, \left( \begin{array}{c}
\bigl(V^\dagger_{\scriptscriptstyle\rm CKM}\bigr)_{kl\,}^{} U_{l,L}^{} \\
D_{k,L}^{} \end{array} \right) , ~~~~~~~
Y_u \,\,=\,\, \frac{\sqrt2}{v}\,V^\dagger_{\scriptscriptstyle\rm CKM}\;
{\rm diag}\bigl(m_u^{},m_c^{},m_t^{}\bigr) \,,
\end{eqnarray}
where in the standard parametrization~\cite{pdg}
\begin{eqnarray} \label{vckm}
V_{\scriptscriptstyle\rm CKM}^{} \,= \left(\!\begin{array}{ccc}
 c_{12\,}^{}c_{13}^{} & s_{12\,}^{}c_{13}^{} & s_{13}^{}\,e^{-i\delta}
\vspace{1pt} \\
-s_{12\,}^{}c_{23}^{}-c_{12\,}^{}s_{23\,}^{}s_{13}^{}\,e^{i\delta} & ~~
 c_{12\,}^{}c_{23}^{}-s_{12\,}^{}s_{23\,}^{}s_{13}^{}\,e^{i\delta} ~~ & s_{23\,}^{}c_{13}^{}
\vspace{1pt} \\
 s_{12\,}^{}s_{23}^{}-c_{12\,}^{}c_{23\,}^{}s_{13}^{}\,e^{i\delta} &
-c_{12\,}^{}s_{23}^{}-s_{12\,}^{}c_{23\,}^{}s_{13}^{}\,e^{i\delta} & c_{23\,}^{}c_{13}^{}
\end{array}\right) ,
\end{eqnarray}
with $\delta$ being the $CP$ violation phase, \,$c_{kl}^{}=\cos\theta_{kl}^{}$,\, and
\,$s_{kl}^{}=\sin\theta_{kl}^{}$.\,
We note that, as a consequence, $\Delta_{u8}$ and $\Delta_{d8}$, whose basic building blocks
are $Y^\dagger_u Y_u^{}$ and $Y^\dagger_d Y_d^{}$, respectively, are all diagonal and
thus will not bring about new flavor- and $CP$-violating interactions.

For the lepton sector, since it is still unknown whether light neutrinos are Dirac or
Majorana particles, we address the two possibilities separately.
In the Dirac case, the $M_\nu$ part is absent from ${\cal L}_{\rm m}$
in Eq.\,(\ref{Lm}), which is therefore, in the MFV language, formally invariant under
the~global group
\,${\rm U}(3)_L\times{\rm U}(3)_\nu\times{\rm U}(3)_E =
G_\ell\times{\rm U}(1)_L\times{\rm U}(1)_\nu\times{\rm U}(1)_E$\, with
\,$G_\ell ={\rm SU}(3)_L\times{\rm SU}(3)_\nu\times{\rm SU}(3)_E$.\,
This means that the three generations of $L_{k,L}$, $\nu_{k,R}$, and $E_{k,R}$ transform as
fundamental representations of SU$(3)_{L,\nu,E}$ in $G_{\ell}$, respectively,
\begin{eqnarray}
L_L^{} \,\to\, V_L^{}L_L^{}~, ~~~~~~~ \nu_R^{} \,\to\, V_\nu^{}\nu_R^{}~, ~~~~~~~
E_R^{} \,\to\, V_E^{}E_R^{}~,
\end{eqnarray}
where \,$V_{L,\nu,E}\in\rm SU(3)$,\, whereas the Yukawa couplings are spurions transforming
according to
\begin{eqnarray}
Y_\nu^{} \,\to\, V_L^{}Y_\nu^{}V^\dagger_\nu ~, ~~~~~~~
Y_e^{} \,\to\, V_L^{}Y_e^{}V^\dagger_E ~.
\end{eqnarray}
We will work in the basis where $Y_e$ is already diagonal,
\begin{eqnarray}
Y_e \,\,=\,\, \frac{\sqrt2}{v}\, {\rm diag}\bigl(m_e^{},m_\mu^{},m_\tau^{}\bigr) ~,
\end{eqnarray}
and the fields $\nu_{k,L}$, $\nu_{k,R}$, $E_{k,L}$, and $E_{k,R}$ refer to the mass eigenstates.
We can then express $L_{k,L}$ and $Y_\nu$ in terms of the Pontecorvo-Maki-Nakagawa-Sakata (PMNS)
neutrino mixing matrix $U_{\scriptscriptstyle\rm PMNS}$~as
\begin{eqnarray}
L_{k,L}^{} \,= \left( \!\begin{array}{c} (U_{\scriptscriptstyle\rm PMNS})_{kl\,}^{}
\nu_{l,L}^{} \vspace{2pt} \\ E_{k,L}^{} \end{array}\! \right) , ~~~~~~~
Y_\nu \,=\, \frac{\sqrt2}{v}\,U_{\scriptscriptstyle\rm PMNS}^{}\,\hat m_\nu^{} ~, ~~~~
\hat m_\nu^{} \,=\, {\rm diag}\bigl(m_1^{},m_2^{},m_3^{}\bigr) ~, \label{Ynu}
\end{eqnarray}
where $m_{1,2,3}^{}$ are the light neutrino eigenmasses and $U_{\scriptscriptstyle\rm PMNS}$
has the same standard parametrization as in Eq.\,(\ref{vckm}).
Thus the discussion for the down-type quarks can be easily applied to the charged leptons by
replacing $V_{\scriptscriptstyle\rm CKM}$ with $U_{\scriptscriptstyle\rm PMNS}^\dagger$ and
employing the building blocks \,${\sf A}=Y_\nu^{}Y^\dagger_\nu$\, and
\,${\sf B}=Y_e^{}Y^\dagger_e$\, to construct $\Delta_\nu$ and $\Delta_e$, which are the lepton
counterparts of $\Delta_u$ and $\Delta_d$, respectively.

If neutrinos are of Majorana nature, the $M_\nu$ part in Eq.\,(\ref{Lm}) is allowed.
As a consequence, for \,$M_\nu\gg M_{\rm D}=v Y_\nu/\sqrt{2}$\, the seesaw
mechanism~\cite{seesaw1} becomes operational involving the 6$\times$6 neutrino mass
matrix
\begin{eqnarray}
{\sf M} \,\,=\,\, \left ( \begin{array}{cc} 0 & M_{\rm D}^{} \vspace{2pt} \\
M_{\rm D}^{\rm T} & M_\nu^{} \end{array} \right)
\end{eqnarray}
in the $\bigl(U_{\scriptscriptstyle\rm PMNS}^*\nu^{\rm c}_L, \nu_R^{}\bigr){}^{\rm T}$ basis.
The resulting matrix of light neutrino masses is
\begin{eqnarray} \label{mnu}
m_\nu \,\,=\,\, -\frac{v^2}{2}\, Y_\nu^{}M_\nu^{-1}Y_\nu^{\rm T} \,\,=\,\,
U_{\scriptscriptstyle\rm PMNS\,}^{}\hat m_{\nu\,}^{}U_{\scriptscriptstyle\rm PMNS}^{\rm T} ~,
\end{eqnarray}
where now $U_{\scriptscriptstyle\rm PMNS}$ contains the diagonal matrix
\,$P={\rm diag}(e^{i\alpha_1/2},e^{i\alpha_2/2},1)$\, multiplied from the right,
$\alpha_{1,2}^{}$ being the Majorana phases.
It follows that $Y_\nu$ in Eq.\,(\ref{Ynu}) is no longer valid, and one can instead take $Y_\nu$
to be~\cite{Casas:2001sr}
\begin{eqnarray} \label{ym}
Y_\nu^{} \,\,=\,\,
\frac{i\sqrt2}{v}\,U_{\scriptscriptstyle\rm PMNS\,}^{}\hat m^{1/2}_\nu O M_\nu^{1/2} ~,
\end{eqnarray}
where $O$ is a matrix satisfying \,$OO^{\rm T}=\openone$\,
and \,$M_\nu={\rm diag}(M_1,M_2,M_3)$.\,
As we will see later, $O$~can provide a potentially important new source of $CP$ violation
besides $U_{\scriptscriptstyle\rm PMNS}$.
We comment that the presence of $M_\nu$ breaks the global U(3)$_\nu$ completely if $M_{1,2,3}$
are unequal and partially into O(3)$_\nu$ if $M_{1,2,3}$ are equal~\cite{Cirigliano:2005ck}.

\section{Fermion EDMs in MFV framework\label{fedm}}

To explore the MFV contribution to the EDMs of quarks and charged leptons, one needs
to construct the relevant operators using $\Delta_{u,d,e}$ in combination with the quark,
lepton, Higgs, and gauge fields.
At leading order, the operators can be written as~\cite{D'Ambrosio:2002ex,Cirigliano:2005ck}
\begin{eqnarray} \label{operators}
O^{(u1)}_{RL} \,=\, g'\bar U_R^{}Y^\dagger_u\Delta_{qu1}^{}\sigma_{\kappa\omega}^{}
\tilde H^\dagger Q_L^{}B^{\kappa\omega} ~, & ~~~~~~~ &
O^{(u2)}_{RL} \,=\, g\bar U_R^{}Y^\dagger_u\Delta_{qu2\,}\sigma_{\kappa\omega}^{}
\tilde H^\dagger\tau_a^{}Q_L^{}W_a^{\kappa\omega} ~,
\nonumber \\
O^{(d1)}_{RL} \,=\, g'\bar D_R^{}Y^\dagger_d\Delta_{qd1}^{}\sigma_{\kappa\omega}^{}
H^\dagger Q_L^{}B^{\kappa\omega} ~, & ~~~~~~~ &
O^{(d2)}_{RL} \,=\, g_{\,}\bar D_R^{}Y^\dagger_d\Delta_{qd2\,}^{}\sigma_{\kappa\omega}^{}
H^\dagger\tau_a^{}Q_L^{}W_a^{\kappa\omega} ~, \vphantom{\int_{\int_|^|}}
\\ \label{operators'}
O^{(e1)}_{RL} \,=\, g'\bar E_R^{}Y^\dagger_e\Delta_{\ell1}^{}\sigma_{\kappa\omega}^{}
H^\dagger L_L^{}B^{\kappa\omega} ~, & ~~~~~~~ &
O^{(e2)}_{RL} \,=\, g_{\,}\bar E_R^{}Y^\dagger_e\Delta_{\ell2\,}^{}\sigma_{\kappa\omega}^{}
H^\dagger\tau_a^{}L_L^{}W_a^{\kappa\omega} ~,
\end{eqnarray}
where $W$ and $B$ denote the usual SU(2)$_L\times{\rm U(1)}_Y$ gauge fields with coupling
constants $g$ and~$g'$, respectively, $\tau_a^{}$ are Pauli matrices, \,$a=1,2,3$\, is summed
over, and $\Delta_{qu\varsigma,qd\varsigma,\ell\varsigma}$ with \,$\varsigma=1,2$\, have
the same form as $\Delta$ in Eq.\,(\ref{general}), but generally different $\xi_r^{}$.
One can express the effective Lagrangian containing these operators as
\begin{eqnarray} \label{Leff}
{\cal L}_{\rm eff} \,\,=\,\, \frac{1}{\Lambda^2} \Bigl(O^{(u1)}_{RL} + O^{(u2)}_{RL}
+ O^{(d1)}_{RL} + O^{(d2)}_{RL} + O^{(e1)}_{RL} + O^{(e2)}_{RL}\Bigr) \;+\; {\rm H.c}. ~,
\end{eqnarray}
where $\Lambda$ is the MFV scale.
In general the operators in ${\cal L}_{\rm eff}$ have their own coefficients which
have been absorbed by $\xi_r^{}$ in their respective $\Delta$'s.
These coefficients also take into account the possibility that the MFV scale in
the quark sector may differ from that in the lepton sector.

Expanding Eq.\,(\ref{Leff}), one can identify the terms relevant to fermion EDMs.
In the quark sector the resulting EDMs of up- and down-type quarks are, respectively,
proportional to
\,Im$\bigl(Y^\dagger_u\Delta_{qu\varsigma}V^\dagger_{\scriptscriptstyle\rm CKM}\bigr){}_{kk}^{}$\,
and \,Im$\bigl(Y^\dagger_d\Delta_{qd\varsigma}\bigr){}_{kk}^{}$.\,
The contributions of $\Delta_{qu\varsigma,qd\varsigma}$ to the EDMs come not only from some of
the products of the $\sf A$ and $\sf B$ matrices therein, but also from the imaginary parts
of~$\xi_r^{}$.
As mentioned earlier, ${\rm Im}_{\,}\xi_r^{}$ are always proportional to
\,$J_\xi\equiv{\rm Im\,Tr}\bigl({\sf A}^2{\sf BAB}^2\bigr)=(i/2)\,{\rm Det}[{\sf A,B}]$,\,
or explicitly
\begin{eqnarray} \label{Jxi}
J_\xi^{} \,\,=\,\,
\frac{-64\bigl(m_u^2-m_c^2\bigr)\bigl(m_c^2-m_t^2\bigr)\bigl(m_t^2-m_u^2\bigr)
\bigl(m_d^2-m_s^2\bigr)\bigl(m_s^2-m_b^2\bigr)\bigl(m_b^2-m_d^2\bigr)}{v^{12}}\, J_q^{} ~,
\end{eqnarray}
where
\,$J_q={\rm Im}\bigl(V_{us}^{}V_{cb}^{}V_{ub}^*V_{cs}^*\bigr)=
c_{12\,}^{}s_{12\,}^{}c_{23\,}^{}s_{23\,}^{}c_{13\,}^2s_{13\,}^{}\sin\delta$\,
is a Jarlskog parameter for $V_{\scriptscriptstyle\rm CKM}$.

Not all of the products of \,${\sf A}=Y_u^{}Y_u^\dagger$\, and \,${\sf B}=Y_d^{}Y_d^\dagger$\,
in $\Delta_{qu\varsigma,qd\varsigma}$ will contribute to quark EDMs.
Since $Y_u$ has the form in Eq.\,(\ref{Yu}) and $Y_d$ is diagonal, the Hermiticity of $\sf A$
and $\sf B$ implies that only certain combinations of them are relevant.
For example, \,$\bigl(Y^\dagger_d{\sf A}\bigr){}_{kk}^{}=\sqrt2\,m_{D_k}{\sf A}_{kk}/v$\, is
purely real and hence does not affect $d_{D_k}$.
In this case, one needs to have terms in $\Delta_{qd\varsigma}$ which are not Hermitian in
order to have imaginary components in $\bigl(Y_d^\dagger\Delta_{qd\varsigma}\bigr){}_{kk}^{}$.
We find that only two terms, proportional to ${\sf B}^2{\sf AB}$ and ${\sf B}^2{\sf A}^2{\sf B}$,
are pertinent to the up-type quarks' EDMs and only the ${\sf ABA}^2$ and ${\sf AB}^2{\sf A}^2$
terms are pertinent to the EDM's of down-type quarks.

The preceding discussions show that the contributions of ${\rm Im}_{\,}\xi_r^{}$
to the EDM of, say, the $u$~($d$) quark
are suppressed by a factor of \,$m^2_c/v^2$ $\bigl(m^2_s m^2_b/v^4\bigr)$ compared to
the contributions from ${\sf B}^2\sf AB$ $\bigl({\sf ABA}^2\bigr)$, which has the least number
of suppressive factor from $Y_u$~$(Y_d)$ among the products in Eq.\,(\ref{general}) that can
potentially contribute.
Hence we can neglect the impact of ${\rm Im}_{\,}\xi_r$ on the quark~EDMs.
One, however, needs to take ${\rm Im}_{\,}\xi_r$ into account when considering
how MFV affects the contribution of the strong theta-term to the neutron~EDM,
as we demonstrate later.

Simplifying things, we arrive at the leading-order contributions to the $u$- and $d$-quarks' EDMs
\begin{eqnarray} \label{du}
d_u &\,=\,& \frac{\sqrt2\,e_{\,}v}{\Lambda^2}\;
{\rm Im}\bigl[Y^\dagger_u\bigl(\Delta_{qu1}+\Delta_{qu2}\bigr)
V^\dagger_{\scriptscriptstyle\rm CKM}\bigr]_{11}
\nonumber \\
&\,=\,& \frac{32_{\,}e_{\,}m_u^{}}{\Lambda^2} \Biggl[ \xi^u_{15} \,+\,
\frac{2\bigl(m_c^2+m^2_t\bigr)}{v^2}\,\xi^u_{17} \Biggr]
\frac{\bigl(m_c^2-m_t^2)(m^2_d - m_s^2)(m_s^2-m_b^2)(m_d^2-m_b^2)}{v^8}\, J_q^{} ~,
\\ \nonumber \\ \label{dd}
d_d &\,=\,& \frac{\sqrt2\,e_{\,}v}{\Lambda^2}\;
{\rm Im}\bigl[Y^\dagger_d\bigl(\Delta_{qd1}-\Delta_{qd2}\bigr)\bigr]_{11}
\nonumber \\
&\,=\,& \frac{32_{\,}e_{\,}m_d^{}}{\Lambda^2} \Biggl[ \xi^d_{12} \,+\,
\frac{2\bigl(m_s^2+m^2_b\bigr)}{v^2}\, \xi^d_{16} \Biggr]
\frac{\bigl(m_s^2-m_b^2\bigr)\bigl(m^2_u-m_c^2\bigr)\bigl(m_c^2-m_t^2\bigr)
\bigl(m_u^2-m_t^2\bigr)}{v^8}\, J_q^{} ~, ~~~
\end{eqnarray}
where \,$\xi^u_r=\xi^{u1}_r+\xi^{u2}_r$\, and \,$\xi^d_r=\xi^{d1}_r-\xi^{d2}_r$.\,
The expressions for $d_{c,t}$ and $d_{s,b}$ can be simply derived from Eqs.\,\,(\ref{du})
and (\ref{dd}), respectively, by cyclically changing the quark labels.\footnote{\baselineskip=12pt
It is worth commenting that since \,${\rm Im}_{\,}\xi_r^{}\propto{\rm Det}[{\sf A,B}]$,\, due to
the reality of the coefficients $\xi_{jkl\cdots\,}^{}$ in the infinite series expansion of $\Delta$,
and since $\sf A$ and $\sf B$ are Hermitian, $d_q$ would be identically zero if there were only one
generation of fermions. The same applies to the lepton sector.}

In the lepton sector, we get from Eq.\,(\ref{Leff}) the electron EDM
\begin{eqnarray} \label{de}
d_e &\,=\,& \frac{\sqrt2\,e_{\,}v}{\Lambda^2}\; {\rm Im}\bigl(Y^\dagger_e \Delta_{\ell1}^{}
\,-\, Y^\dagger_e \Delta_{\ell2}^{}\bigr)_{11}
\nonumber \\ &\,=\,&
\frac{\sqrt2\,e_{\,}v}{\Lambda^2} \Bigl[ \xi^\ell_{12}\,
{\rm Im}\bigl(Y^\dagger_e {\sf ABA}^2\bigr)_{11} +
\xi^\ell_{16}\,{\rm Im}\bigl(Y^\dagger_e {\sf AB}^2{\sf A}^2\bigr)_{11} \Bigr] \,, ~~
\end{eqnarray}
where \,$\xi^\ell_r=\xi^{\ell1}_r-\xi^{\ell2}_r$,\, we have ignored ${\rm Im}_{\,}\xi^\ell_r$,
and here \,${\sf A}=Y_\nu^{}Y_\nu^\dagger$\, and \,${\sf B}=Y_e^{}Y_e^\dagger$.\,
If neutrinos are Dirac particles, analogously to $d_d$, we obtain
\begin{eqnarray} \label{dedirac}
d_e^{\rm D} \,=\, \frac{32 e_{\,}m_e^{}}{\Lambda^2} \Biggl[ \xi^\ell_{12} +
\frac{2\bigl(m_\mu^2+m^2_\tau\bigr)}{v^2}\,\xi^\ell_{16} \Biggr]
\frac{\bigl(m_\mu^2-m_\tau^2\bigr)\bigl(m^2_1-m_2^2\bigr)\bigl(m_2^2-m_3^2\bigr)
\bigl(m_3^2-m_1^2\bigr)}{v^8}\, J_\ell^{} ~, ~
\end{eqnarray}
where \,$J_\ell={\rm Im}\bigl(U_{e2}^{}U_{\mu 3}^{}U_{e3}^*U^*_{\mu2}\bigr)$\, is a Jarlskog
invariant for $U_{\scriptscriptstyle\rm PMNS}$.

In the case of Majorana neutrinos, if $\nu_{k,R}^{}$ are degenerate, \,$M_\nu={\cal M}\openone$,\,
and $O$ is a~real orthogonal matrix,\footnote{\baselineskip=12pt Since the lepton Lagrangian with
$\nu_{k,R}^{}$ being degenerate is O(3)$_\nu$ symmetric, one could transform this real $O$
into a unit matrix~\cite{Cirigliano:2006nu}.}
from Eq.\,(\ref{ym}) we have
\begin{eqnarray}
{\sf A} \,\,=\,\, \frac{2}{v^2}\, {\cal M}_{\,}U_{\scriptscriptstyle\rm PMNS\,}^{}\hat m^{}_\nu
U_{\scriptscriptstyle\rm PMNS}^\dagger
\end{eqnarray}
and consequently
\begin{eqnarray} \label{dem}
d_e^{\rm M} \,\,=\,\, \frac{32e_{\,}m_{e\,}^{}{\cal M}^3}{\Lambda^2 v^8}
\bigl(m_\mu^2-m_\tau^2\bigr) \bigl(m_1^{}-m_2^{}\bigr)\bigl(m_2^{}-m_3^{}\bigr)
\bigl(m_3^{}-m_1^{}\bigr)\,\xi_{12\,}^\ell J_\ell^{} ~,
\end{eqnarray}
the $\xi^\ell_{16}$ term having been neglected.
Since \,$m_k^{}\ll\cal M$,\, we can see that $d_e^{\rm D}$ is highly suppressed relative
to $d_e^{\rm M}$.
The formulas for $d_{\mu,\tau}^{\rm D}$ and $d_{\mu,\tau}^{\rm M}$ can be readily found from
Eqs.\,\,(\ref{dedirac}) and (\ref{dem}), respectively, by cyclically changing
the mass subscripts.

In the discussion above, $d_e$ arises from the $CP$-violating Dirac phase $\delta$ in
$U_{\scriptscriptstyle\rm PMNS}$, and the Majorana phases $\alpha_{1,2}^{}$ therein do not participate.
However, if $\nu_{k,R}^{}$ are not degenerate, nonzero $\alpha_{1,2}^{}$ can bring about an additional
effect on $d_e$, even with a real \,$O\neq\openone$.\,
With a complex $O$, the phases in it may give rise to an extra contribution to $d_e$, whether or
not $\nu_{k,R}^{}$ are degenerate.
The formulas for $d_e$ in these scenarios are more complicated than Eq.\,(\ref{dem}) and are not
shown here, but we will explore some of them numerically in the next section.

The various contributions to the fermion EDMs that we have considered have high powers in
Yukawa couplings.
Since the MFV hypothesis presupposes that all $CP$-violation effects originate from
the Yukawa couplings, the high orders in them reflect the fact that nonvanishing EDMs in
the SM begin to appear at the three-loop level for quarks and in higher loops for the electron.
One may wonder whether these are the only contributions to fermion EDMs under the MFV framework.
The answer is no because one can realize fermion EDMs by combining some lower-order Yukawa terms
from the MFV operators with SM loop diagrams, such as those contributing to quark EDMs in the~SM.
Nevertheless, hereafter we will not include such type of possible contributions.
The contributions that we have already covered should provide a good idea about how fermion EDMs
are generated in the presence of MFV.
For definiteness, we will apply numerically the results we have acquired and discuss some of
their implications.

\section{Numerical analysis\label{numerics}}

We will first treat the neutron EDM, $d_n$, evaluated from the quark
contributions and infer from its data a bound on the scale of quark MFV.
We will also look at how MFV affects the contribution of the strong $\theta$-term to~$d_n$.
Proceeding to the lepton sector, we will devote much of the section to the electron EDM, and
briefly deal with the muon and tau EDMs,
in order to explore limitations on the scale of leptonic MFV.
Afterwards, we will examine constraints from $CP$-violating electron-nucleon interactions
which were probed by recent searches for atomic and molecular EDMs.
Finally, we will address potential restrictions from some $CP$-conserving processes.

\subsection{Neutron EDM\label{nedm}}

In calculating quark EDMs, as in Eqs.\,\,(\ref{du}) and (\ref{dd}),
one needs to take into account the running of the quark masses due to QCD evolution.
We adopt the mass ranges \,$m_u^{}=0.00139_{-0.00041}^{+0.00042}$,\,
$m_d^{}=0.00285_{-0.00048}^{+0.00049}$,\, $m_s^{}=0.058_{-0.012}^{+0.018}$,\,
$m_c^{}=0.645_{-0.085}^{+0.043}$,\, $m_b^{}=2.90_{-0.06}^{+0.16}$,\, and
\,$m_t^{}=174.2\pm1.2$,\, all in  GeV, at a~renormalization scale \,$\mu=m_W^{}$\,
from Ref.$\;$\cite{Xing:2011aa}.
With the central values of these masses and the quark Jarlskog parameter
\,$J_q=\bigl(3.02_{-0.19}^{+0.16}\bigr)\times10^{-5}$\, from the latest fit by
CKMfitter~\cite{ckmfit}, we arrive at
\begin{eqnarray} \label{dq}
d_u^{} &\,=\,& \frac{1.4\times10^{-35\;}e\,\rm cm}{\Lambda^2/\rm GeV^2}
\bigl(\xi_{15}^u+\xi_{17}^u\bigr) ~, ~~~~~~~~~
d_d^{} \,\,=\,\, \frac{1.3\times10^{-29\;}e\,\rm cm}{\Lambda^2/\rm GeV^2}
\bigl(\xi_{12}^d+0.00028\;\xi_{16}^d\bigr) ~, ~~~~
\nonumber \\
d_s^{} &\,=\,& \frac{-2.6\times10^{-28\;}e\,\rm cm}{\Lambda^2/\rm GeV^2}
\bigl(\xi_{12}^d+0.00028\;\xi_{16}^d\bigr) ~,
\end{eqnarray}
where \,$\xi_r^u=\xi_r^{u1}+\xi_r^{u2}$\, and \,$\xi_r^d=\xi_r^{d1}-\xi_r^{d2}$.\,
Evidently, the $s$-quark effect may be dominant.

To determine the neutron EDM, one needs to connect it to the quark-level quantities.
The relation between $d_n$ and $d_{u,d,s}$ can be parameterized as
\begin{eqnarray}
d_n^{} \,\,=\,\, \eta_n^{}
\bigl(\rho_{n\,}^u d_u^{}+\rho_{n\,}^d d_d^{}+\rho_{n\,}^s d_s^{}\bigr) \,,
\end{eqnarray}
where \,$\eta_n^{}=0.4$\, accounts for corrections due to the QCD evolution from
\,$\mu=m_W^{}$\, down to the hadronic scale~\cite{Degrassi:2005zd} and the values of
the parameters $\rho_n^{u,d,s}$ depend on the model for the matrix elements
\,$\langle n|\bar q\sigma^{\kappa\omega}q|n\rangle=\rho_{n\,}^q\bar u_n^{}\sigma^{\kappa\omega}u_n^{}$.\,
For instance, in the constituent quark model
\,$\rho_n^d=\frac{4}{3}=-4\rho_n^u$\, and \,$\rho_n^s=0$\,~\cite{He:1989xj}, whereas in the parton quark
model \,$\rho_n^u=-0.508$, \,$\rho_n^d=0.746$,\, and \,$\rho_n^s=-0.226$\,~\cite{Engel:2013lsa}.
From the various models proposed in the literature~\cite{He:1989xj,Engel:2013lsa,Dib:2006hk},
we may conclude that
\begin{eqnarray} \label{rhonq}
-0.78 \,\,\le\,\, \rho_n^u \,\,\le\,\, -0.17 ~, ~~~~~~~
 0.7  \,\,\le\,\, \rho_n^d \,\,\le\,\,  2.1  ~, ~~~~~~~
\mbox{$-0.35$} \,\,\le\,\, \rho_n^s \,\,\le\,\, 0 ~.
\end{eqnarray}
In view of these numbers and Eq.\,(\ref{dq}), we can ignore the $d_u$ and $\xi_{16}^d$ terms.
Hence, taking the extreme values \,$\rho_n^d=2.1$\, and \,$\rho_n^s=-0.35$,\, as well as
scanning over the quark mass and $J_q$ ranges quoted above to maximize~$d_n$, we get
\begin{eqnarray} \label{dnmax}
d_n^{} \,\,=\,\, \frac{8.4\times10^{-29\;}e\,\rm cm}{\Lambda^2/\rm GeV^2}\;\xi_{12}^d ~.
\end{eqnarray}
It is then interesting to note that
\,$\Lambda/\bigl|\xi_{12}^d\bigr|\raisebox{1pt}{$^{1/2}$}=100$\,GeV\, translates into
\,$d_n^{}=8.4\times10^{-33\;}e$\,cm,\, which is roughly similar to the SM expectation
\,$d_n^{\scriptscriptstyle\rm SM}\sim10^{-32}$-$10^{-31\;}e$\,cm\,~\cite{hesm}.
Comparing Eq.\,(\ref{dnmax}) with the current data \,$|d_n|_{\rm exp}<2.9\times10^{-26\;}e$\,cm\,
at 90\% CL~\cite{pdg}, we extract
\begin{eqnarray}
\frac{\Lambda}{\bigl|\xi_{12}^d\bigr|\raisebox{1pt}{$^{1/2}$}} \,\,>\,\, 0.054\;\rm GeV ~,
\end{eqnarray}
which is not strict at all.
Less extreme choices of $\rho_n^{d,s}$ would lead to even weaker bounds.
We conclude that the present neutron-EDM limit cannot yield a useful restriction on $\Lambda$.

One can also look at the contributions of quark chromo-EDMs to the neutron EDM~\cite{He:1989xj}.
The relevant operators are obtainable from the MFV quark-EDM operators by replacing
$W_a^{\mu\nu}$ and $\tau_a^{}$ with the gluon field strength tensor $G_c^{\mu\nu}$ and
the color SU(3) generators $\lambda_c^{}$, respectively.
The extracted constraints on $\Lambda$ are similar.

\subsection{MFV contribution to strong theta term\label{thetaterm}}

Besides the quark (chromo-)EDMs, another contributor to the neutron EDM is the theta term
of QCD~\cite{'tHooft:1976up}, which in the SM is given by~\cite{Engel:2013lsa}
\begin{eqnarray}
{\cal L}_{\bar\theta}^{} \,\,=\,\, \frac{-\bar\theta\!\;g_{\rm s}^2}{32\pi^2}\,
\epsilon_{\kappa\upsilon\phi\omega}^{}G_c^{\kappa\upsilon}G_c^{\phi\omega} ~,
\end{eqnarray}
where \,$\bar\theta=\theta+\arg{\rm Det}(Y_u Y_d)$\, involves the bare $\theta$-parameter,
$g_{\rm s}^{}$ is the strong coupling constant, and~\,$\epsilon_{0123}^{}=+1$.\,
The inclusion of MFV causes $\bar\theta$ to be modified to
\begin{eqnarray} \label{thetamfv}
\bar\theta_{\scriptscriptstyle\rm MFV} \,\,=\,\, \theta \,+\,
\arg{\rm Det}\bigl(\Delta_{qu}^\dagger Y_{u\,}^{}\Delta_{qd}^\dagger Y_d^{}\bigr)
\,\,=\,\, \bar\theta \,+\,
\arg{\rm Det}_{\,}\Delta_{qu}^\dagger+\arg{\rm Det}_{\,}\Delta_{qd}^\dagger ~,
\end{eqnarray}
where $\Delta_{qu,qd}$ have the same expression as $\Delta$ in Eq.\,(\ref{general}),
but generally different coefficients $\xi_r^{}$.
Although the addition of these new factors to the Yukawa Lagrangian amounts only to
a redefinition of $Y_{u,d}$ and hence has no direct experimental implications after
the quark mass matrices are diagonalized, we can expect that $\Delta_{qu,qd}$ are
close to the unit matrix.
Our interest is in investigating the size of \,$\arg{\rm Det}_{\,}\Delta_{qu,qd}$
in Eq.\,(\ref{thetamfv}) and thus whether or not their presence makes the fine tuning
between the two terms in $\bar\theta$ worse.

To compute ${\rm Det}_{\,}\Delta_{qu}$, we first write the real and imaginary parts of $\xi_r^{}$
in terms of real constants $\varrho_r^{}$ and $\imath_r^{}$ as
\begin{eqnarray}
{\rm Re}_{\,}\xi_r^{} \,\,=\,\, \varrho_r^{} ~, ~~~~~~~
{\rm Im}_{\,}\xi_r^{} \,\,=\,\, \imath_r^{}\,J_\xi^{}
\end{eqnarray}
with $J_\xi$ given in Eq.\,(\ref{Jxi}).
Upon applying the Cayley-Hamilton identity, we then get
\begin{eqnarray}
{\rm Det}_{\,}\Delta_{qu} \,\,=\,\,
\mbox{$\frac{1}{6}$}\bigl({\rm Tr}\Delta_{qu}\bigr)^3
- \mbox{$\frac{1}{2}$}\, {\rm Tr}\Delta_{qu}\, {\rm Tr}\bigl(\Delta_{qu}^2\bigr)
+ \mbox{$\frac{1}{3}$}\,{\rm Tr}\bigl(\Delta_{qu}^3\bigr) ~,
\end{eqnarray}
which leads us to
\begin{eqnarray} \label{redelta}
{\rm Re}\bigl({\rm Det}_{\,}\Delta_{qu}\bigr) &\,\simeq&\, \varrho_1^3 \,+\,
\varrho_{1\,}^2 \bigl( \varrho_{2\,}^{} y_t^2 + \varrho_{4\,}^{} y_t^4 \bigr) ~,
\vphantom{|_{\int_\int^\int}^{}} \\ \label{imdelta}
J_\xi^{-1}\,{\rm Im}\bigl({\rm Det}_{\,}\Delta_{qu}\bigr) &\,\simeq&\,
-\varrho_{2\,}^{} \bigl[ \varrho_{2\,}^{}\varrho_{15}^{} +
\varrho_{3\,}^{} \bigl(\varrho_{13}^{}-\varrho_{14}^{}\bigr) +
\varrho_{5\,}^{}\varrho_9^{} + \varrho_{7\,}^{}\varrho_{11}^{} \bigr]
- \varrho_{3\,}^{} \bigl( \varrho_{3\,}^{}\varrho_{12}^{}-\varrho_{4\,}^{}\varrho_{11}^{}
- \varrho_{6\,}^{}\varrho_9^{} \bigr)
\nonumber \\ && -\;
\bigl(\varrho_6^{}-\varrho_7^{}\bigr) \bigl( \varrho_{2\,}^{}\varrho_{10}^{}
+ \varrho_{3\,}^{}\varrho_8^{}+\varrho_{4\,}^{}\varrho_5^{}-\varrho_{6\,}^{}\varrho_7^{} \bigr)
\,-\, \varrho_{2\,}^{} \bigl( \varrho_{4\,}^{}\varrho_{17}^{} +
\varrho_{9\,}^{}\varrho_{13}^{} \bigr) y_t^4
\nonumber \\ && +\;
\bigl[ -\varrho_{2\,}^{} \bigl( \varrho_{2\,}^{}\varrho_{17}^{} +
\varrho_{4\,}^{}\varrho_{15}^{} - \varrho_{6\,}^{}\varrho_{14}^{} +
\varrho_{7\,}^{}\varrho_{13}^{} + \varrho_{9\,}^{}\varrho_{11}^{} \bigr)
- \varrho_{3\,}^{} \bigl(\varrho_{6\,}^{}\varrho_{12}^{}-\varrho_{8\,}^{}\varrho_9^{}\bigr)
\nonumber \\ && ~~~~ -
\bigl(\varrho_6^{}-\varrho_7^{}\bigr) \bigl( \varrho_{4\,}^{}\varrho_{10}^{}
- \varrho_{6\,}^{}\varrho_9^{} \bigr) \bigr] y_t^2
\nonumber \\ && +\;
\varrho_1^{~\,} \bigl\{ -\varrho_{2\,}^{}\varrho_{17}^{} - \varrho_{3\,}^{}\varrho_{16}^{} +
\varrho_{4\,}^{}\varrho_{15}^{} + \varrho_{5\,}^{}\varrho_{12}^{} +
\varrho_{6\,}^{}\varrho_{13}^{} - \varrho_{7\,}^{}\varrho_{14}^{} +
\varrho_{8\,}^{}\varrho_{11}^{} - \varrho_{9\,}^{}\varrho_{10}^{}
\nonumber \\ && ~~~~~~~ +
\bigl[ 2_{\,}\varrho_{2\,}^{}\imath_1^{}-\varrho_{6\,}^{}\varrho_{16}^{} +
\varrho_{8\,}^{}\bigl(\varrho_{13}^{}-\varrho_{14}^{}\bigr) +
\varrho_{11\,}^{}\varrho_{12}^{} \bigr] y_t^2
\nonumber \\ && ~~~~~~~ +
\bigl(2_{\,}\varrho_{4\,}^{}\imath_1^{}+\varrho_{12\,}^{}\varrho_{13}^{}\bigr)y_t^4 \bigr\}
\nonumber \\ && +~
\varrho_{1\,}^2 \bigl( 3_{\,}\imath_1^{}+\imath_{2\,}^{} y_t^2+\imath_{4\,}^{} y_t^4 \bigr) ~,
\end{eqnarray}
where \,$y_q^{}=\sqrt2\, m_q^{}/v$\, and on the right-hand sides we have ignored terms
suppressed by powers of~$y_{u,c,d,s,b}$.
The formulas for ${\rm Det}_{\,}\Delta_{qd}$ are similar.

Since \,$y_t^2\sim1\gg y_{u,c,d,s,b}^2$,\, the requirement that
\,$\Delta_{qu,qd}\simeq\openone$\, implies that
\begin{eqnarray}
\varrho_1^{} \,\,\simeq\,\,1 ~, ~~~~~~~ \bigl|\varrho_{2,4}^{}\bigr|\,\,\ll\,\,1 ~, ~~~~~~~
\bigl|\varrho_{3,5,6,\ldots,17}^{}\bigr|\,\,\le\,\, {\cal O}(1) ~, ~~~~~~~
\bigl|\imath_{1,2,\ldots,17}^{}\bigr|\,\,\le\,\, {\cal O}(1) ~,
\end{eqnarray}
Using these conditions and the quark parameter values employed earlier, we have checked
numerically that Eqs.\,\,(\ref{redelta}) and (\ref{imdelta}) approximate well the exact
(but much lengthier) expressions, especially if~\,$|\varrho_{2,4}^{}|\le{\cal O}(0.001)$.\,
Moreover, we get \,$|{\rm arg}\,{\rm Det}_{\,}\Delta_{qu,qd}|<10^{-21}$.\,
Obviously, the MFV effect is negligible compared to the present bound
\,$\bar\theta_{\rm exp}<10^{-10}$~\cite{Engel:2013lsa}.

\subsection{Electron EDM\label{eedm}}

To evaluate the EDMs of charged leptons, we need the values of
the various pertinent quantities, such as the elements of the neutrino mixing matrix
$U_{\scriptscriptstyle\rm PMNS}$ as well as the masses of neutrinos and charged leptons.
If neutrinos are Dirac in nature, the parametrization of $U_{\scriptscriptstyle\rm PMNS}$ is
the same as $V_{\scriptscriptstyle\rm CKM}$ in Eq.\,(\ref{vckm}).
In Table\,\,\ref{nudata}, we have listed $\sin^2\!\theta_{kl}$ and $\delta$ from a recent fit
to global neutrino data~\cite{Capozzi:2013csa}.  Most of these numbers depend on whether
neutrino masses fall into a~normal hierarchy~(NH), where \,$m_1^{}<m_2^{}<m_3^{}$,\,
or an inverted one (IH), where \,$m_3^{}<m_1^{}<m_2^{}$.\,
If neutrinos are Majorana particles, $U_{\scriptscriptstyle\rm PMNS}$ contains an additional
matrix \,$P={\rm diag}(e^{i\alpha_1/2},e^{i\alpha_2/2},1)$\, multiplied from the right,
where $\alpha_{1,2}^{}$ are the Majorana phases which remain unknown.

Also listed in Table\,\,\ref{nudata} are the differences in neutrinos' squared masses,
which are well determined.
In contrast, our knowledge about the absolute scale of the masses is still poor.
Some information on the latter is available from tritium $\beta$-decay
experiments~\cite{nureview}.  In particular, their latest results imply an upper limit on
the (electron based) antineutrino mass of \,$m_{\bar\nu_e}<2$\,eV~\cite{pdg}.
Planned measurements will be more sensitive by an order of magnitude~\cite{nureview}.
Indirectly, stronger bounds on the total mass \,$\Sigma_k^{}m_k^{}=m_1^{}+m_2^{}+m_3^{}$\,
can be inferred from cosmological observations.
Specifically, the Planck Collaboration extracted \,$\Sigma_k^{}m_k^{}<0.23$\,eV\, at 95\%\,CL
from cosmic microwave background (CMB) and baryon acoustic oscillation (BAO)
measurements~\cite{planck}.  Including additional observations can
improve this limit to \,$\Sigma_k^{}m_k^{}<0.18$\,eV~\cite{Riemer-Sorensen:2013jsa}.
On the other hand, there are also recent analyses that have turned up tentative indications
of bigger masses and hence quasidegeneracy (QD) among the neutrinos.
The South Pole Telescope Collaboration reported \,$\Sigma_k^{}m_k^{}=(0.32\pm0.11)$\,eV\,
from the combined CMB, BAO, Hubble constant, and Sunyaev-Zeldovich selected galaxy cluster
abundances dataset~\cite{Hou:2012xq}.
This is compatible with the later finding \,$\Sigma_k^{}m_k^{}=(0.36\pm0.10)$\,eV\, favored by
the Baryon Oscillation Spectroscopic Survey CMASS Data Release 11~\cite{Beutler:2014yhv}.
In the following numerical work, we take this QD possibility into consideration.

\begin{table}[b]
\caption{Results of a recent fit to the global data on neutrino oscillations~\cite{Capozzi:2013csa}.
The neutrino mass hierarchy may be normal $\bigl(m_1^{}<m_2^{}<m_3^{}\bigr)$ or inverted
$\bigl(m_3^{}<m_1^{}<m_2^{}\bigr)$.\label{nudata}} \small
\begin{tabular}{|c||c|c|} \hline Observable & NH & IH \\ \hline\hline
$\sin^2\theta_{12}^{}$ &
$0.308\pm0.017\vphantom{\frac{1}{2}_|^|}$ & $0.308\pm0.017\vphantom{\frac{1}{2}_|^|}$
\\
$\sin^2\theta_{23}^{}$ & $0.425_{-0.027}^{+0.029}$        &
$0.437_{-0.029}^{+0.059}\vphantom{\frac{1}{2}_|^|}$
\\
$\sin^2\theta_{13}^{}$ & $0.0234_{-0.0018}^{+0.0022}$      &
$0.0239\pm0.0021\vphantom{\frac{1}{2}_|^|}$
\\
$\delta/\pi$ & $1.39_{-0.27}^{+0.33}$ & $1.35_{-0.39}^{+0.24}\vphantom{\frac{1}{2}_|^|}$
\\
$\Delta m_{21}^2 = m_2^2-m_1^2$    &
$\left(7.54_{-0.22}^{+0.26}\right)_{\vphantom{\int}}^{\vphantom{\int}}\times10^{-5}\;\rm eV^2$ &
$\left(7.54_{-0.22}^{+0.26}\right)\times10^{-5}\;\rm eV^2$
\\
~$\Delta m^2 = \bigl|m_3^2-\bigl(m_1^2+m_2^2\bigr)/2\bigr|\vphantom{\int_{|_|}^{|^|}}$~ &
~$\bigl(2.44_{-0.06}^{+0.08}\bigr)\times10^{-3}\;\rm eV^2$~ &
~$(2.40\pm0.07)\times10^{-3}\;\rm eV^2$~
\\ \hline
\end{tabular}
\end{table}

If neutrinos are of Dirac nature, we first note that the mass difference definitions in
Table\,\,\ref{nudata} imply that
\,$\bigl(m^2_1-m_2^2\bigr)\bigl(m_2^2-m_3^2\bigr)\bigl(m_3^2-m_1^2\bigr)=
\Delta m_{21}^2\bigl(\Delta m^2\bigr)\raisebox{1pt}{$^2$}
- \frac{1}{4}\bigl(\Delta m_{21}^2\bigr)\raisebox{1pt}{$^3$}$,\,
which is independent of $m_k^{}$ individually.
Then, scanning the parameter ranges in Table\,\,\ref{nudata} to maximize~$d_e^{\rm D}$
in Eq.\,(\ref{dedirac}), we obtain for the NH (IH) of neutrino masses~\cite{He:2014fva}
\begin{eqnarray} & \displaystyle
d_e^{\rm D} \,\,=\,\, \frac{1.3\;(1.3)\times10^{-99\;}e\,\rm cm}{\Lambda^2/\rm GeV^2}\;
\xi_{12}^\ell ~, & \label{ded}
\end{eqnarray}
after dropping the $\xi_{16}^\ell$ part.
This is negligible compared to the latest data \,$|d_e|_{\rm exp}<8.7\times10^{-29\;}e$\,cm\,
reported by ACME~\cite{acme}, and the smallness is due to the light neutrino masses being tiny.

In contrast, if neutrinos are Majorana particles, $d_e$ can be sizable.
To see this, we begin with the simplest possibility that $\nu_{k,R}^{}$ are degenerate,
\,$M_\nu={\cal M}\openone$,\, and the $O$ matrix in Eq.\,(\ref{ym}) is real.
For this scenario, $d_e$ is already given in Eq.\,(\ref{dem}),
which depends on the choice for one of $m_{1,2,3}^{}$ after the mass data are included.
Scanning again the empirical parameter ranges in Table\,\,\ref{nudata} to
maximize~$d_e^{\rm M}$, we obtain for \,$m_1^{}=0$\, $\bigl(m_3^{}=0\bigr)$ in the NH (IH) case
\begin{eqnarray} \label{deM'}
\!\!\frac{d_e^{\rm M}}{e\,\rm cm} \,\,=\,\, 4.7\;(0.52)\times10^{-23}\biggl(\frac{{\cal M}}
{10^{15\,}\rm GeV}\biggr)^{\!3}\biggl(\frac{\rm GeV}{\hat\Lambda}\biggr)^{\!2} ,
\end{eqnarray}
where \,$\hat\Lambda=\Lambda/\bigl|\xi_{12}^\ell\bigr|\raisebox{1pt}{$^{1/2}$}$.\,
Then \,$|d_e^{\rm exp}|<8.7\times10^{-29}\;e$\,cm~\cite{acme} implies
\begin{eqnarray} \label{lambda}
\hat\Lambda \,\,>\,\, 0.74\;(0.24){\rm\;TeV}\;
\biggl(\frac{{\cal M}}{10^{15\,}\rm GeV}\biggr)^{\!3/2} .
\end{eqnarray}

Although this might suggest that $\hat\Lambda$ could be extremely high with
an excessively large ${\cal M}$, there are limitations on ${\cal M}$.
Since the series in Eq.\,(\ref{general}), which implicitly incorporates arbitrarily high powers
of $\sf A$ and~$\sf B$, has to converge, their eigenvalues need to be
capped~\cite{Mercolli:2009ns,He:2014fva}.
Otherwise, the coefficients $\xi_r^{}$ might not converge to finite numbers after
the reduction of $\Delta$ from its infinite series expansion to Eq.\,(\ref{general}).
In the lepton sector, we only need to be concerned with \,${\sf A}=Y_\nu^{}Y_\nu^\dagger$,\,
as \,${\sf B}=Y_e^{}Y_e^\dagger$\, already has diminished eigenvalues.
Thus one may demand that the eigenvalues of $\sf A$ are at most~1.
However, since MFV may emerge from calculations of SM loops, the expansion quantities may be
more naturally be ${\sf A}/(16\pi^2)$ and ${\sf B}/(16\pi^2)$, in which case the maximum
eigenvalue of $\sf A$ cannot be more than~$16\pi^2$.
As another alternative, one may impose the perturbativity condition on the Yukawa
couplings, namely \,$(Y_\nu)_{jk}<\sqrt{4\pi}$\,~\cite{Kanemura:1999xf}, implying a cap of
$4\pi$ instead.

In this paper we require the eigenvalues of \,${\sf A}=Y_\nu^{}Y_\nu^\dagger$\,
not to exceed unity.
Furthermore, in our illustrations we will choose the largest eigenmasses of
the right-handed neutrinos subject to this condition.
For the example resulting in Eq.\,(\ref{lambda}), this translates into the maximal value
\,${\cal M}=6.16\;(6.22)\times 10^{14}\,$GeV\, in the NH (IH) case and consequently
\begin{eqnarray} \label{hatLambda0}
\hat\Lambda \,\,>\,\, 0.36\;(0.12){\rm\;TeV} ~.
\end{eqnarray}
This constraint would weaken if \,$m_{1(3)}>0$.\,
For comparison with later illustrations, the $\cal M$ numbers above translate into
\,$d_e^{\rm M}\hat\Lambda^2=1.1\;(0.13)\times10^{-23}\,e$\,cm.\,

Now, with $\nu_{k,R}^{}$ still degenerate, \,$M_\nu={\cal M}\openone$,\, but $O$ complex, $\sf A$
has a less simple expression,
\begin{eqnarray}
{\sf A} \,\,=\, \frac{2}{v^2}\,{\cal M}_{\,}U_{\scriptscriptstyle\rm PMNS\,}^{}\hat m^{1/2}_\nu
OO^\dagger\hat m^{1/2}_\nu U_{\scriptscriptstyle\rm PMNS}^\dagger ~,
\end{eqnarray}
which is to be applied to $d_e^{\rm M}$ in Eq.\,(\ref{de}).
From now on, we ignore the $\xi_{16}^\ell$ parts.
We can always write \,$OO^\dagger=e^{2i\sf R}$\, with a~real antisymmetric matrix
\begin{eqnarray}
{\sf R} \,\,=\, \left ( \begin{array}{ccc} 0 & r_1^{} & r_2^{} \vspace{1pt} \\
-r_1^{} & 0 & r_3^{} \vspace{1pt} \\ -r_2^{} & ~ \mbox{$-r_3^{}$} ~ & 0 \end{array} \right) .
\end{eqnarray}
Since $OO^\dagger$ is not diagonal, $\sf A$ will in general have dependence on
the Majorana phases in $U_{\scriptscriptstyle\rm PMNS}$ if they are not zero.
To concentrate first on demonstrating how $O$ can give rise to $CP$ violation beyond
that induced by the Dirac phase~$\delta$ in $U_{\scriptscriptstyle\rm PMNS}$, we switch off
the Majorana phases,~\,$\alpha_{1,2}^{}=0$.\,
Subsequently, for illustrations, we pick two possible sets of~$r_{1,2,3}^{}$, namely,
(i)~$r_1^{}=-r_2^{}=r_3^{}=-\rho$\, and (ii)~$r_1^{}=2r_2^{}=3r_3^{}=\rho$,\,
and employ the central values of the data in Table\,\,\ref{nudata}, particularly
\begin{eqnarray} \label{delta}
\delta \,\,=\,\, 1.39\pi ~{\rm(NH)~~or~~} 1.35\pi ~\rm (IH) ~.
\end{eqnarray}

\begin{figure}[b]
\includegraphics[height=43mm,width=73mm]{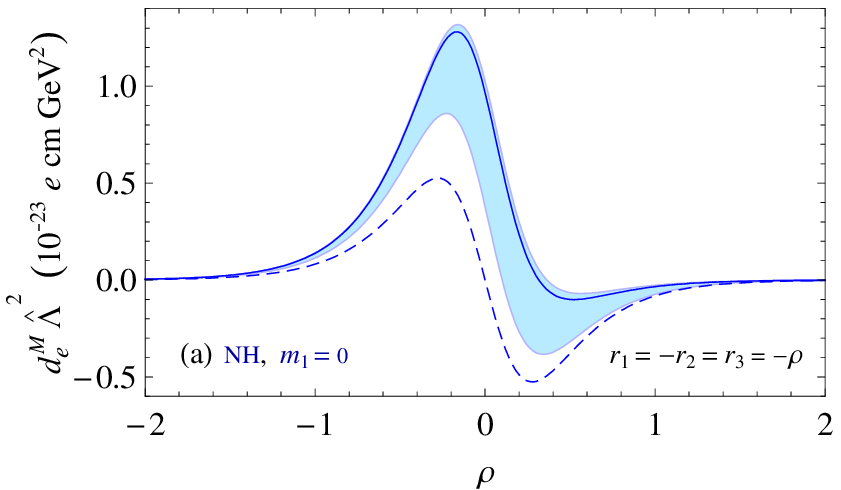} ~ ~
\includegraphics[height=43mm,width=203pt]{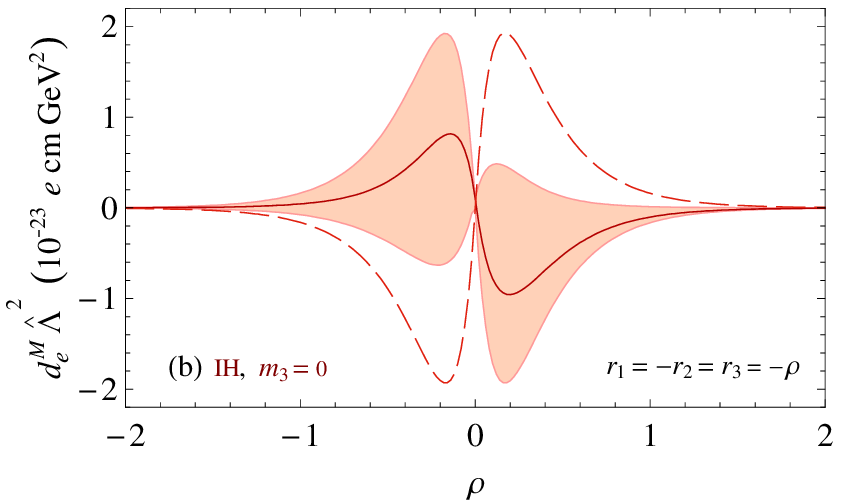}\smallskip\\
\includegraphics[height=43mm,width=73mm]{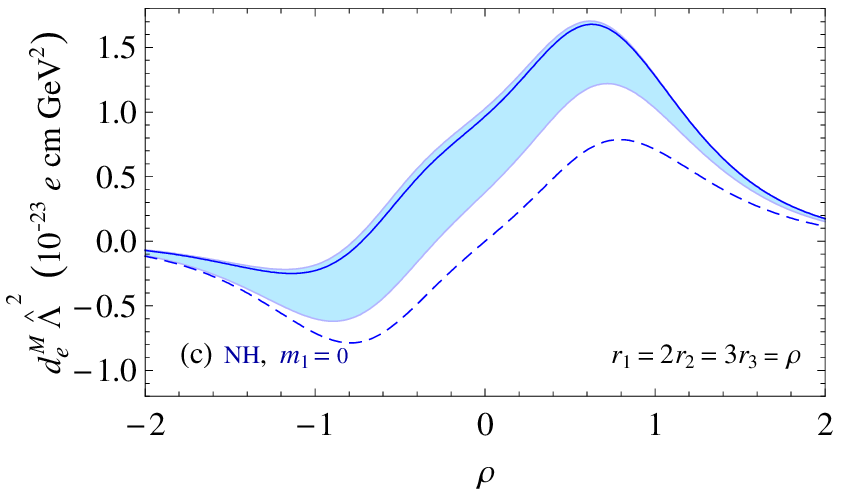} ~ ~
\includegraphics[height=43mm,width=203pt]{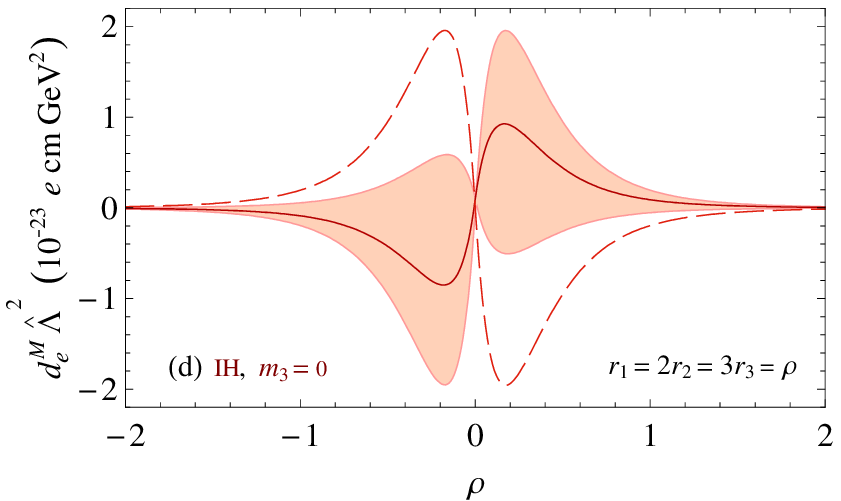}\smallskip\\
~ \includegraphics[height=43mm,width=71mm]{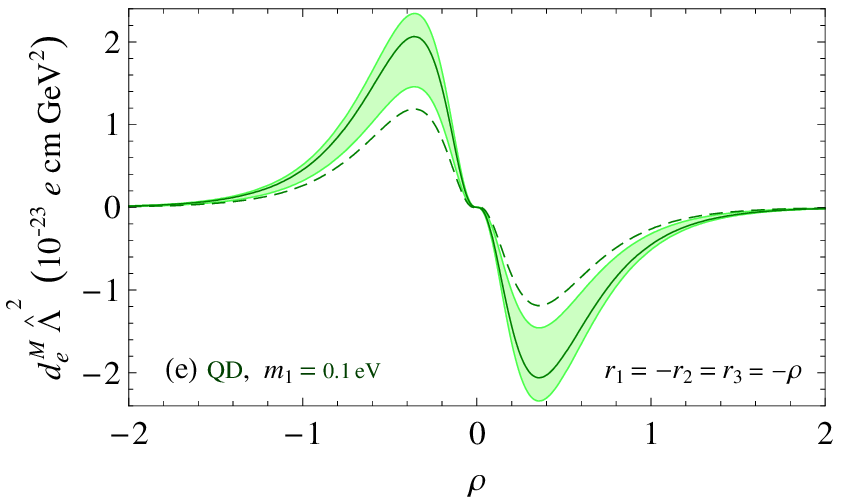} ~
\includegraphics[height=43mm,width=210pt]{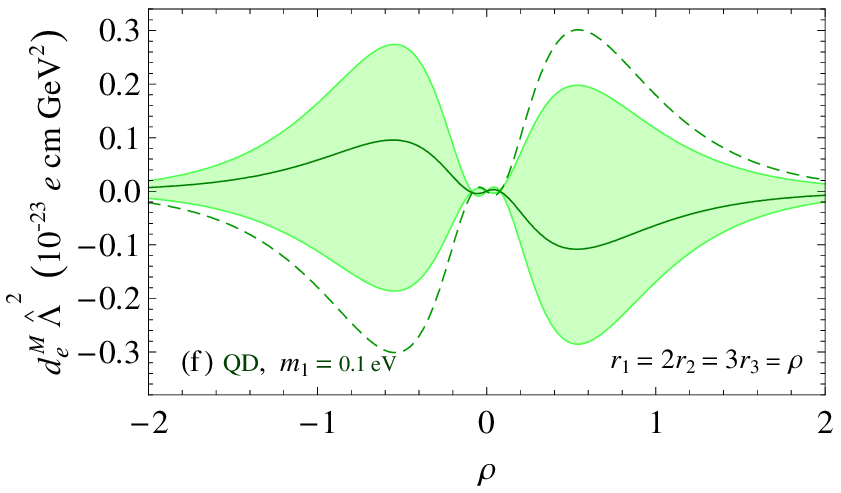}\vspace{-7pt}
\caption{Dependence of $d_e^{\rm M}$ times $\hat\Lambda^2=\Lambda^2/\bigl|\xi_{12}^\ell\bigr|$\,
on the $O$-matrix parameter $\rho$ in the absence of the Majorana phases, \,$\alpha_{1,2}^{}=0$,\,
for degenerate $\nu_{k,R}^{}$ and complex $O$ with (a,b,e)~$r_1^{}=-r_2^{}=r_3^{}=-\rho$\, and
(c,d,f)~$r_1^{}=2r_2^{}=3r_3^{}=\rho$,\, as explained in the text.
The lighter blue, red, and green bands reflect the one-sigma ranges of $\delta$, while the solid
and dashed curves correspond, respectively, to its central values in Eq.\,(\ref{delta}) and
to \,$\delta=0$.
In (e,f) and other QD plots below, only the NH scenario is assumed, unless stated
otherwise.\label{de-rho-complo}}   \vspace{-1ex}
\end{figure}

We present in Fig.$\;$\ref{de-rho-complo}(a)-(d) the resulting $d_e^{\rm M}\hat\Lambda^2$
versus~$\rho$ for the NH (IH) of light neutrino masses with \,$m_{1(3)}=0$.\,
Since $\delta$ is not yet well-determined, we also depict the variations of $d_e^{\rm M}$
over the one-sigma ranges of $\delta$ quoted in Table$\;$\ref{nudata} with the lighter blue
and red bands.  We remark that the boundaries of the bands do not necessarily correspond to
the upper or lower ends of the $\delta$ ranges.
Within these bands, the blue and red solid curves belong, respectively, to the NH and IH
central values in Eq.\,(\ref{delta}).  We also graph the (dashed) curves for \,$\delta=0$\,
to reveal the $CP$-violating role of $O$ alone.
The solid and dashed curves in Fig.$\;$\ref{de-rho-complo}(a,b) are roughly the mirror
images about \,$\rho=0$\, of the corresponding curves given in Ref.\,\cite{He:2014fva} for
\,$r_{1,2,3}^{}=\rho$.\,

In Fig.$\;$\ref{de-rho-complo}(a)-(d), as well as in Ref.\,\cite{He:2014fva}, we have only
examples where the lightest neutrinos are massless and, consequently, the neutrino masses
sum up to \,$\Sigma_k^{}m_k^{}=0.059\;$eV and 0.099$\;$eV\, in the NH and IH cases, respectively.
These numbers satisfy the aforementioned bound from cosmological data,
\,$\Sigma_k^{}m_k^{}<0.18$\,eV\,~\cite{Riemer-Sorensen:2013jsa}.
In light of the hints of quasidegenerate neutrinos with \,$\Sigma_k^{}m_k^{}\sim0.3$\,eV\,
from other cosmological observations~\cite{Hou:2012xq,Beutler:2014yhv}, which still need
confirmation by future measurements, here we also provide a couple of instances in
Fig.$\;$\ref{de-rho-complo}(e,f) after making the NH choice
\,$m_1^{}=0.1{\rm\,eV}<m_2^{}<m_3^{}$,\, which translates into \,$\Sigma_k^{}m_k^{}=0.31$\,eV.\,

All these examples in Fig.\,\,\ref{de-rho-complo} clearly indicate that $O$ can generate
potentially significant new effects of $CP$ violation which can exceed those of~$\delta$.
The latter point is most noticeable in Fig.$\;$\ref{de-rho-complo}(b,d) from comparing
the IH \,$\delta\neq0$\, regions at \,$\rho\sim0$\, with the extreme values of
the corresponding IH \,$\delta=0$\, curves.

\begin{figure}[b]
~\includegraphics[width=77mm]{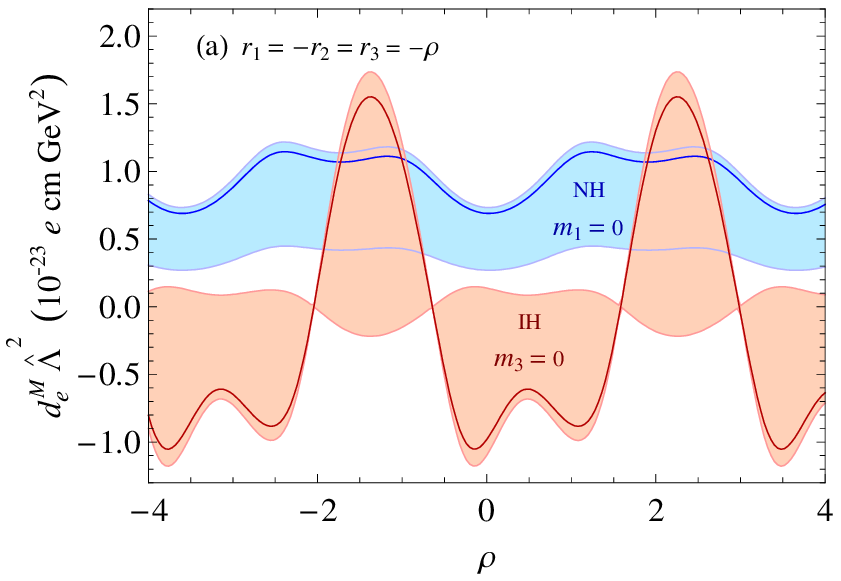} \hspace{7pt}
\includegraphics[width=77mm]{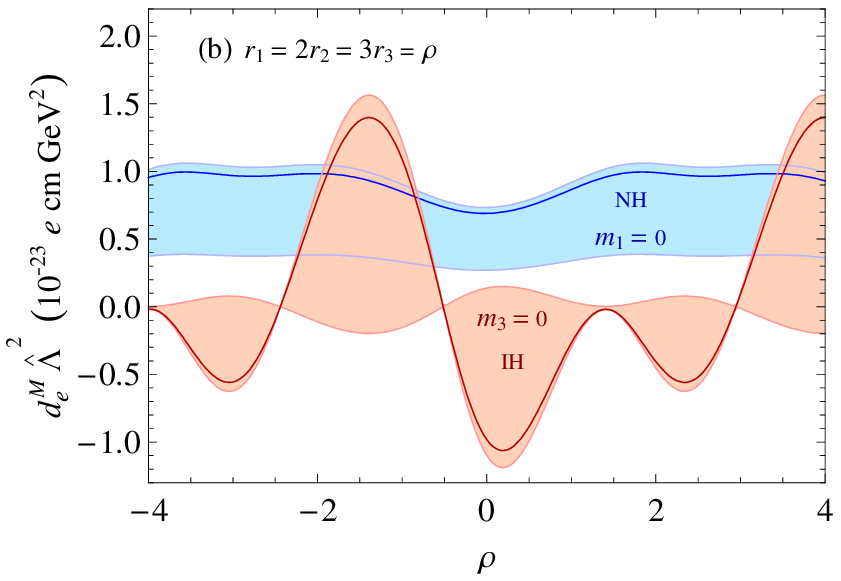}\smallskip\\
\includegraphics[width=224pt]{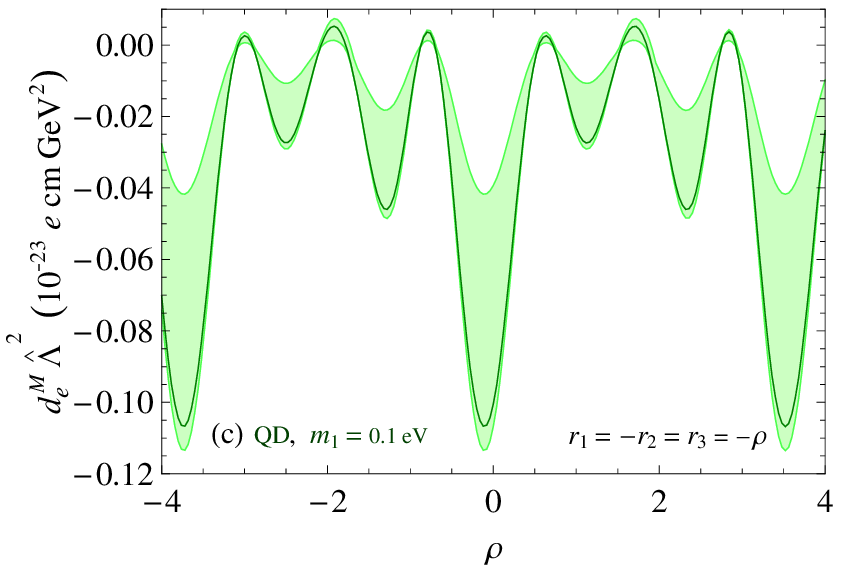} \hspace{1pt}
\includegraphics[width=224pt]{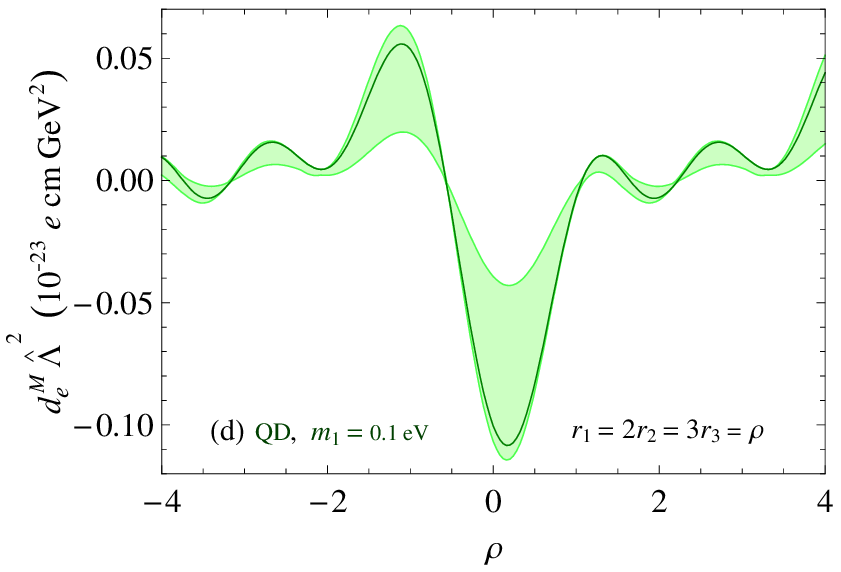}\vspace{-7pt}
\caption{Dependence of \,$d_e^{\rm M}\hat\Lambda^2$\, on $O$-matrix parameter $\rho$ in
the absence of Majorana phases, \,$\alpha_{1,2}^{}=0$,\, for nondegenerate
$\nu_{k,R}^{}$ with \,$M_\nu={\cal M}\,{\rm diag}(1,0.8,1.2)$\, and real \,$O=e^{\sf R}$\, with
(a,c)~$r_1^{}=-r_2^{}=r_3^{}=-\rho$\, and (b,d)~$r_1^{}=2r_2^{}=3r_3^{}=\rho$,\,
as explained in the text.
The lighter blue, red, and green bands reflect the one-sigma ranges of $\delta$, while the solid
curves correspond to its central values in Eq.\,(\ref{delta}).\label{de-rho-realo}}
\end{figure}

With \,$\alpha_{1,2}^{}=0$,\, the $CP$-violating impact of $O$ can still materialize even if it is
real provided that $\nu_{k,R}^{}$ are not degenerate.
In that case
\begin{eqnarray}
{\sf A} \,\,=\, \frac{2}{v^2}\,U_{\scriptscriptstyle\rm PMNS\,}^{}\hat m^{1/2}_\nu O M_\nu
O^\dagger\hat m^{1/2}_\nu U_{\scriptscriptstyle\rm PMNS}^\dagger
\end{eqnarray}
based on Eq.\,(\ref{ym}).
For instance, assuming that $O$ is real, $O=e^{\sf R}$ with \,$r_1^{}=-r_2^{}=r_3^{}=-\rho$,\,
and that \,$M_\nu={\cal M}\,{\rm diag}(1,0.8,1.2)$,\, we show the resulting
\,$d_e^{\rm M}\hat\Lambda^2$\, versus $\rho$ in Fig.$\;$\ref{de-rho-realo}(a), where only
the \,$\delta\neq0$\, curves are nonvanishing and the sinusoidal behavior of $d_e$ is visible.
As in the previous figure, we also display the variations of $d_e^{\rm M}$ over the one-sigma
ranges of $\delta$ from Table$\;$\ref{nudata}.
The solid curves in Fig.$\;$\ref{de-rho-realo}(a) are similar to their \,$r_{1,2,3}^{}=\rho$\,
counterparts in Ref.\,\cite{He:2014fva}.
As another example, we select again \,$r_1^{}=2r_2^{}=3r_3^{}=\rho$,\, keeping the other input
parameters unchanged, and plot Fig.\,\,\ref{de-rho-realo}(b) which differs somewhat
qualitatively from~Fig.\,\,\ref{de-rho-realo}(a).
In Fig.\,\,\ref{de-rho-realo}(c,d) we graph the QD cases with
\,$m_1^{}=0.1{\rm\,eV}<m_2^{}<m_3^{}$, which turn out to have much smaller $d_e^{\rm M}$ ranges.
All of these results further demonstrate the importance of $O$ as an extra source of
$CP$ violation.

\begin{figure}[b]
\includegraphics[width=77mm]{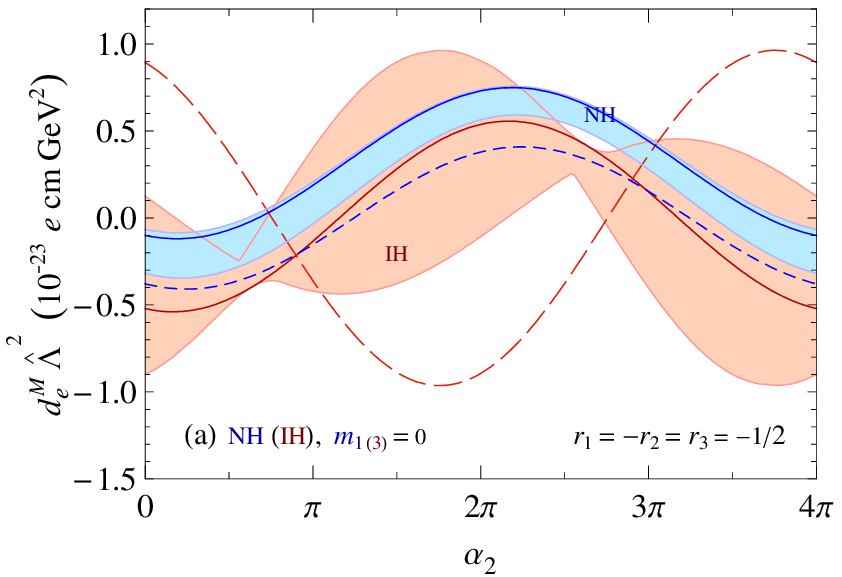}
\includegraphics[width=77mm]{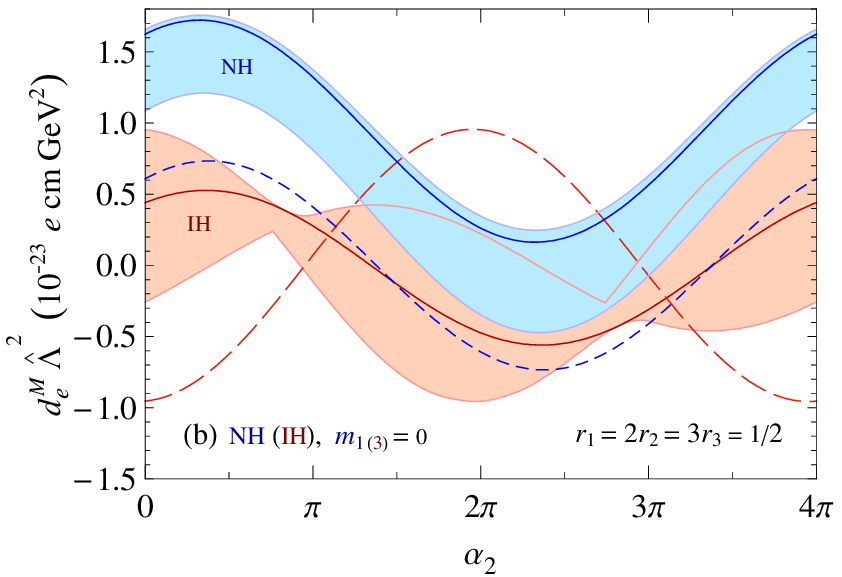}\vspace{-2ex}\\
\includegraphics[width=77mm]{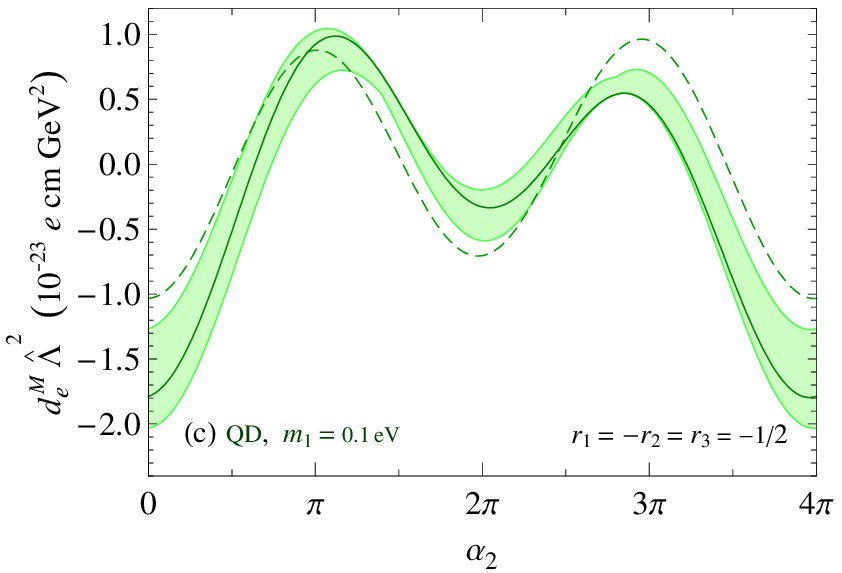} \,
\includegraphics[width=213pt]{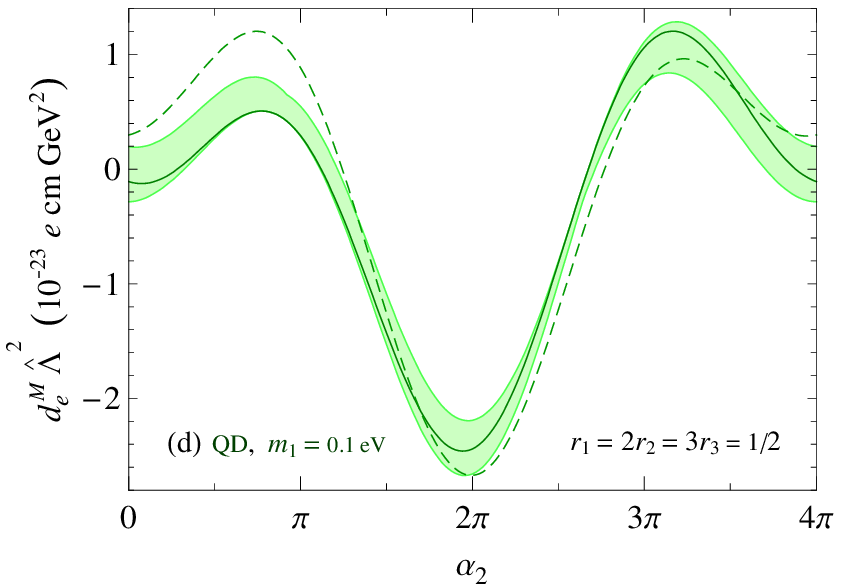}\vspace{-7pt}
\caption{Dependence of \,$d_e^{\rm M}\hat\Lambda^2$\, on $\alpha_2^{}$ for
\,$\alpha_1^{}=0$,\, degenerate $\nu_{k,R}^{}$, and \,$O=e^{i\sf R}$\, with
(a,c)~$r_{1,3}^{}=-r_2^{}=-\frac{1}{2}$\, and (b,d)~$r_1^{}=2r_2^{}=3r_3^{}=\frac{1}{2}$,\,
as explained in the text.  The bands and curves have the same meanings as in preceding
figures.\label{de-alpha-complo}} \vspace{-3ex}
\end{figure}

\begin{figure}[ht]
\includegraphics[width=77mm]{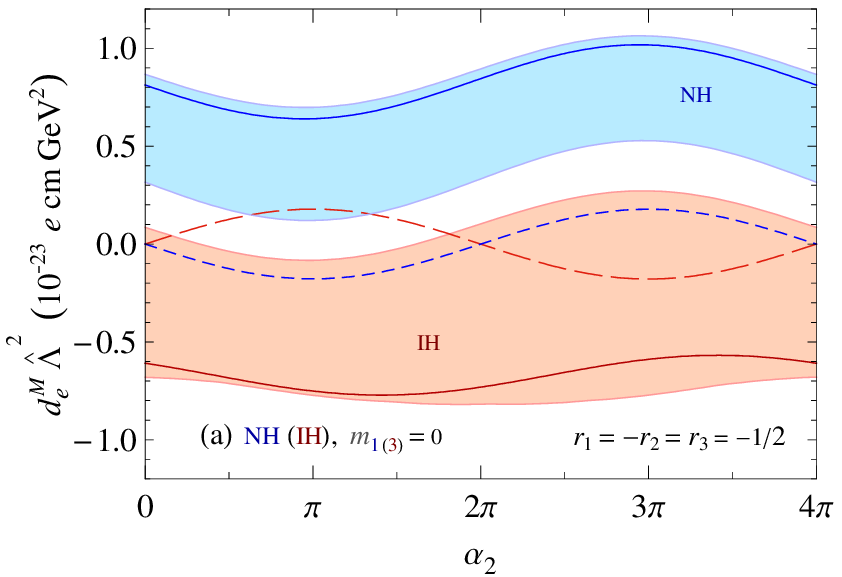}
\includegraphics[width=77mm]{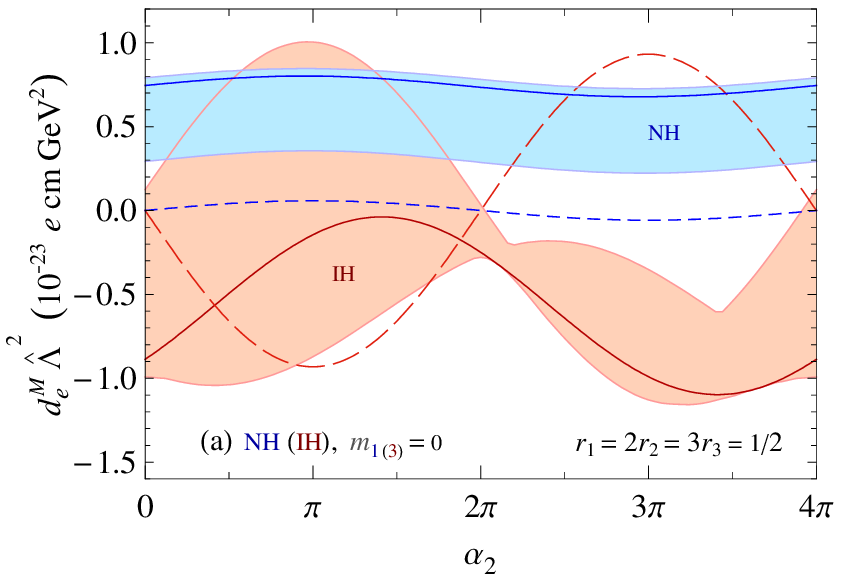}\vspace{-2ex}\\
\includegraphics[width=77mm]{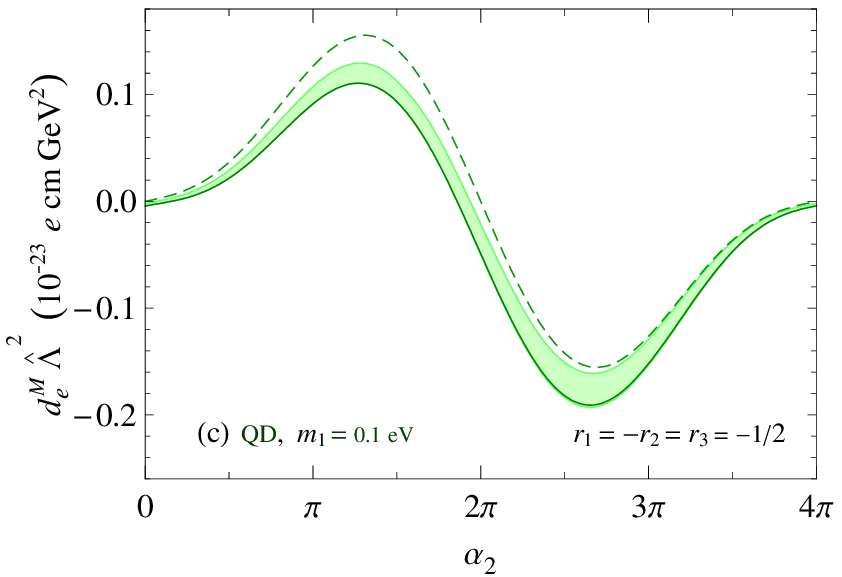}
\includegraphics[width=77mm]{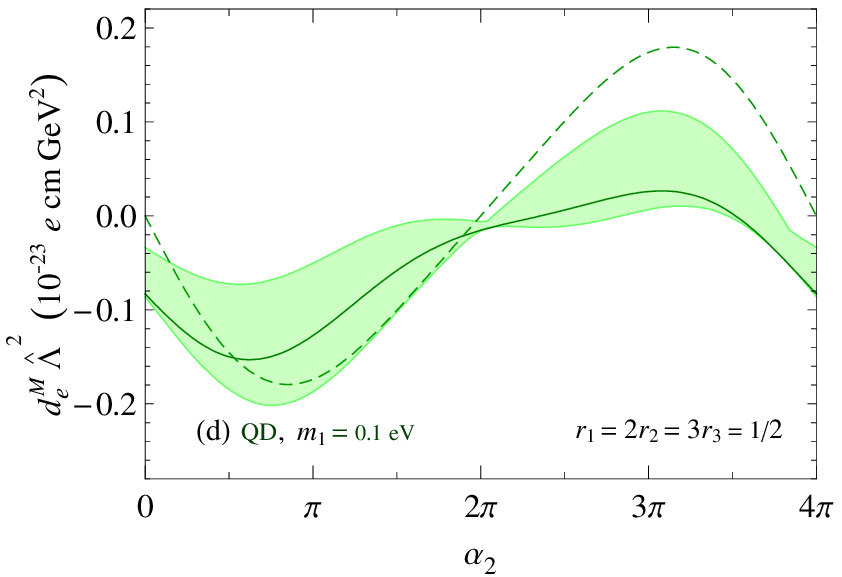}\vspace{-7pt}
\caption{The same as Fig.\,\,\ref{de-alpha-complo}, except $\nu_{k,R}^{}$
are nondegenerate with \,$M_\nu={\cal M}\,{\rm diag}(1,0.8,1.2)$\,
and \,$O=e^{\sf R}$.\label{de-alpha-realo}}
\end{figure}

Turning our attention now to the contribution of the Majorana phases, we first illustrate it
for \,$M_\nu={\cal M}\openone$\, and \,$O=e^{i\sf R}$ with the two sets of $r_{1,2,3}^{}$
chosen in the previous paragraph.
Thus, fixing \,$\alpha_1^{}=0$\, and \,$\rho=\frac{1}{2}$,\, we depict the resulting dependence
of $d_e^{\rm M}$ on $\alpha_2^{}$ in Fig.\,\,\ref{de-alpha-complo} for nonzero $\delta$ within
its one-sigma ranges from Table$\;$\ref{nudata} and also for \,$\delta=0$.\,
For further illustrations, we do the same with \,$M_\nu={\cal M}\,{\rm diag}(1,0.8,1.2)$\,
and \,$O=e^{\sf R}$,\, displaying the results in Fig.\,\,\ref{de-alpha-realo}.
It is noticeable that each of the solid or dashed curves in Figs.~\ref{de-alpha-complo} and
\ref{de-alpha-realo} repeats itself after $\alpha_2^{}$ changes by $4\pi$, which is attributable
to the $e^{i\alpha_2/2}$ dependence of $d_e^{\rm M}$ in these cases.
Also, one can verify visually that the solid curves in Figs.$\;$\ref{de-rho-complo} and
\ref{de-alpha-complo} (\ref{de-rho-realo} and \ref{de-alpha-realo}) are consistent with each
other at \,$\rho=\frac{1}{2}$\, and \,$\alpha_{1,2}^{}=0$.\,
It is evident from the instances in Figs.$\;$\ref{de-alpha-complo} and \ref{de-alpha-realo},
as well as their counterparts in~Ref.\,\cite{He:2014fva}, that the Majorana phases yield
additional important $CP$-violating effects on $d_e$ beyond~$\delta$.

\begin{figure}[t]
\includegraphics[width=57mm]{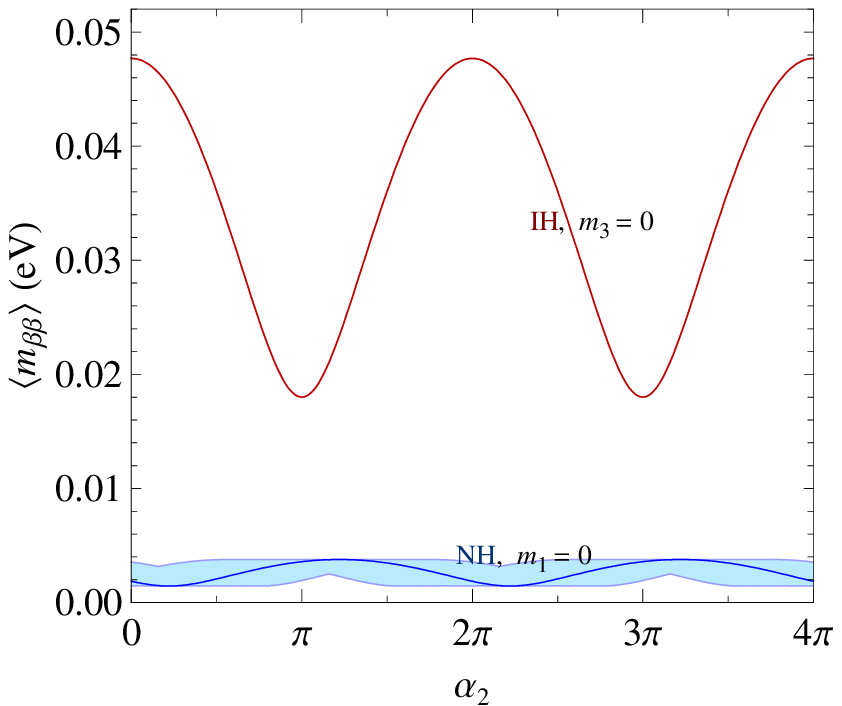}
\includegraphics[width=57mm]{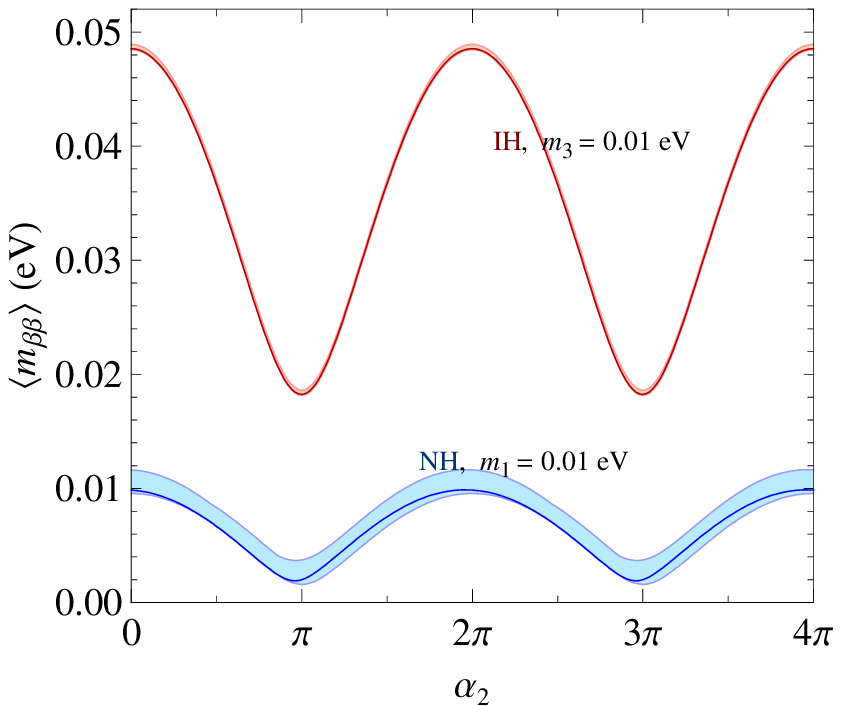}
\includegraphics[width=57mm]{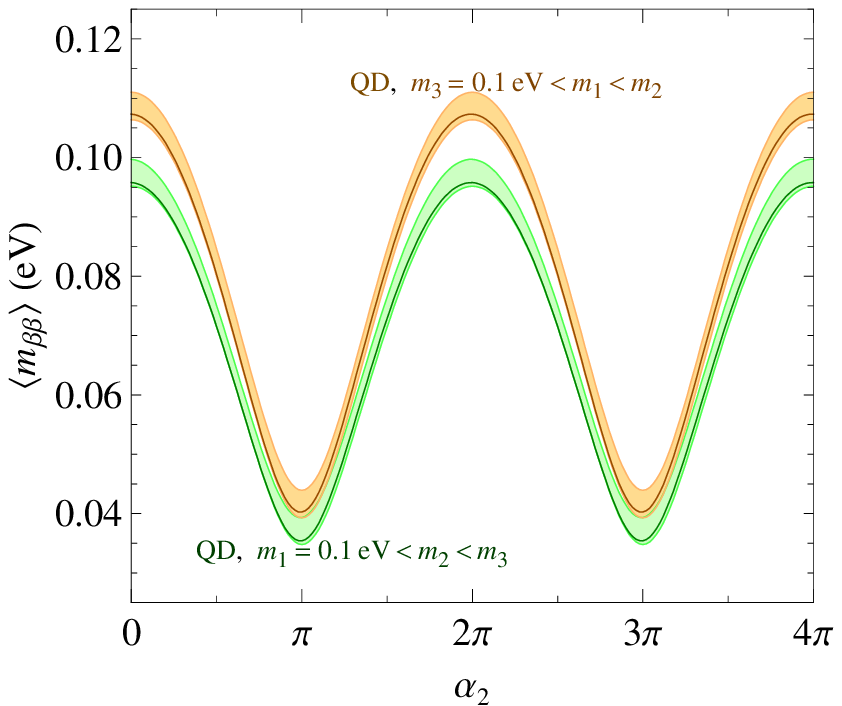}\vspace{-7pt}
\caption{Dependence of effective Majorana mass $\langle m_{\beta\beta}\rangle$ on $\alpha_2^{}$
for \,$\alpha_1^{}=0$,\, nonzero $\delta$, and some selections of $m_{1\rm\;or\;3}$.
The bands and solid curves have the same meanings as in previous figures.\label{mbetabeta}}
\end{figure}

It is interesting that some of the $CP$-violating variables which enter $d_e^{\rm M}$ also
affect neutrinoless double-$\beta$ decay due to the Majorana nature of the electron neutrino.
This process is of fundamental importance because it does not conserve lepton number and thus will
be evidence for new physics if detected~\cite{nureview}.
If there are no other contributions, the rate of neutrinoless double-$\beta$ decay increases with
the square of the effective Majorana mass
\begin{eqnarray} \label{mee}
\bigl\langle m_{\beta\beta}\bigr\rangle &\,=\,& \Bigl|
\raisebox{3pt}{\footnotesize$\displaystyle\sum_k$}\,U_{e k\,}^2\hat m_k^{} \Bigr|
\,\,=\,\, \Bigl|
\left( U_{\scriptscriptstyle\rm PMNS\,}^{}\hat m_\nu^{} U_{\scriptscriptstyle\rm PMNS}^{\rm T}
\right)_{11}\! \Bigl|
\nonumber \\ &\,=\,&
\Bigl| c_{12\,}^2 c_{13\,}^2 m_1^{}\,e^{i\alpha_1} + s_{12\,}^2 c_{13\,}^2 m_2^{}\,e^{i\alpha_2}
+ s_{13\,}^2 m_3^{}\,e^{-2i\delta} \Bigr| ~.
\end{eqnarray}
In Fig.$\;$\ref{mbetabeta} we display several examples of $\langle m_{\beta\beta}\rangle$
versus $\alpha_2^{}$ for \,$\alpha_1^{}=0$,\, but not those for \,$\delta=0$\,
to avoid crowding the plots.
It is obvious that each of the curves repeats itself after $\alpha_2^{}$ changes by $2\pi$,
which is due to the presence of $e^{i\alpha_2}$ in $\langle m_{\beta\beta}\rangle$,
unlike the $d_e^{\rm M}$ curves in Figs.~\ref{de-alpha-complo} and~\ref{de-alpha-realo}.
The peak values in the third plot of Fig.$\;$\ref{mbetabeta} are already close to the existing
experimental upper limits on~$\langle m_{\beta\beta}\rangle$, the best one
being \,0.12\,eV~\cite{xmbb}.  Thus the QD possibility will be tested by forthcoming searches
within the next decade, which are expected to have sensitivities reaching
\,0.04\,eV to 0.01\,eV~\cite{Vignati:2014cqa}.

\begin{figure}[b]
\includegraphics[height=137pt,width=167pt]{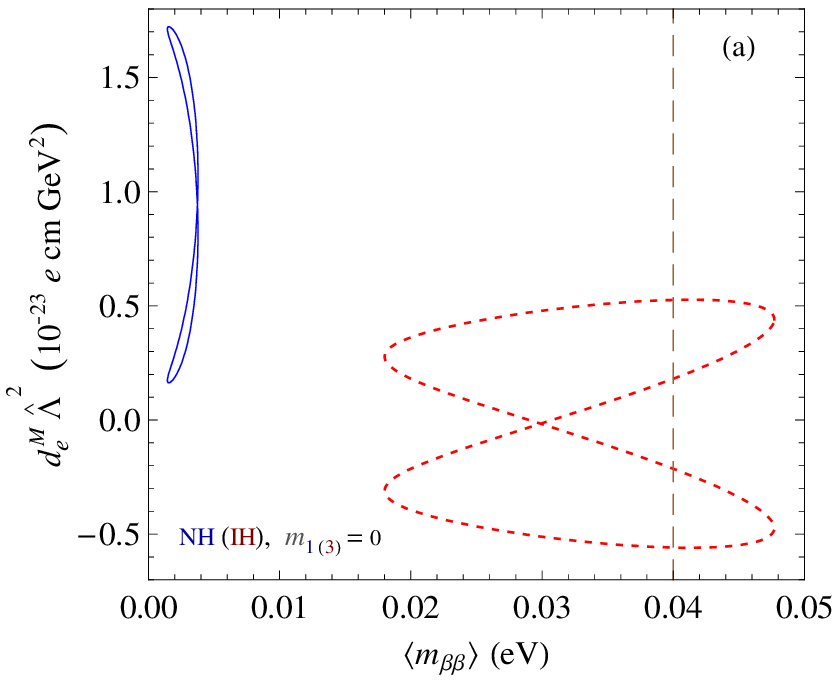} \hspace{-7pt}
\includegraphics[height=137pt,width=163pt]{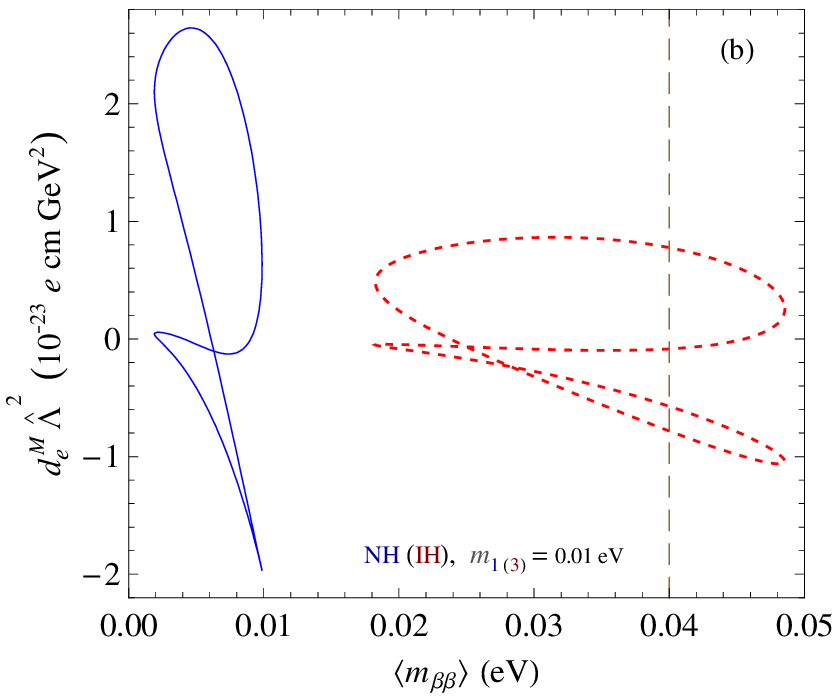} \hspace{-5pt}
\includegraphics[height=137pt,width=157pt]{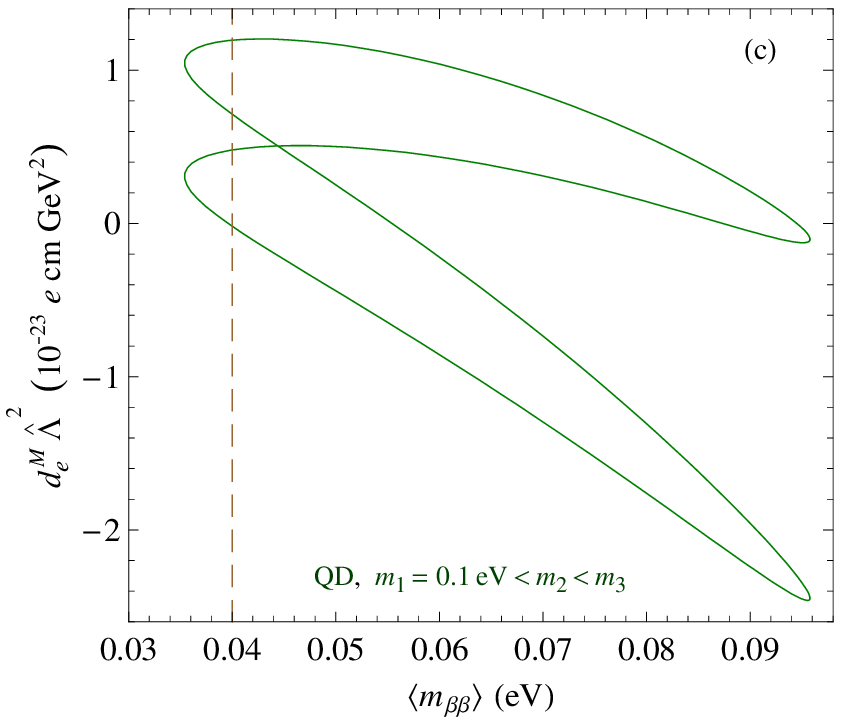} \\
\includegraphics[height=137pt,width=167pt]{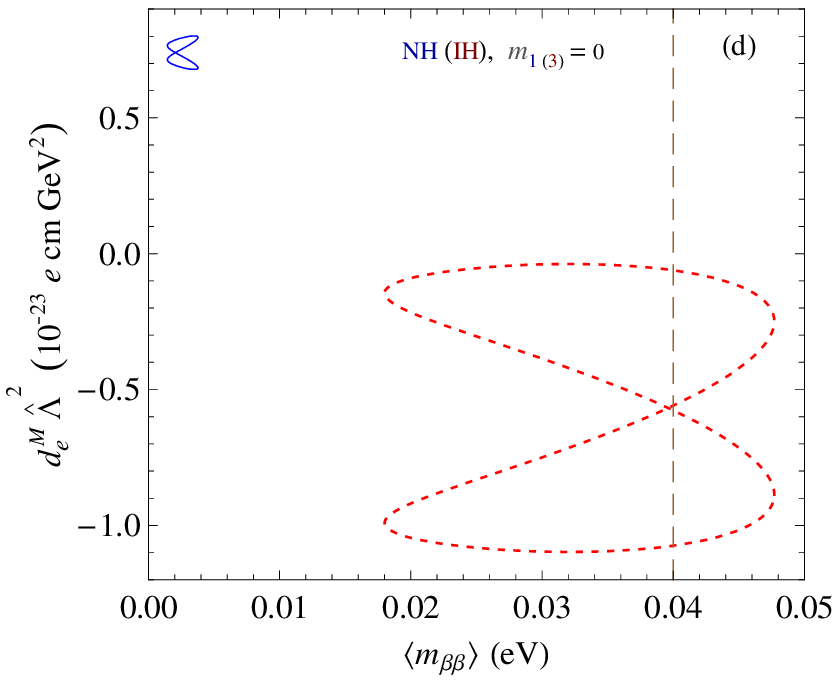} \hspace{-12pt}
\includegraphics[height=137pt,width=168pt]{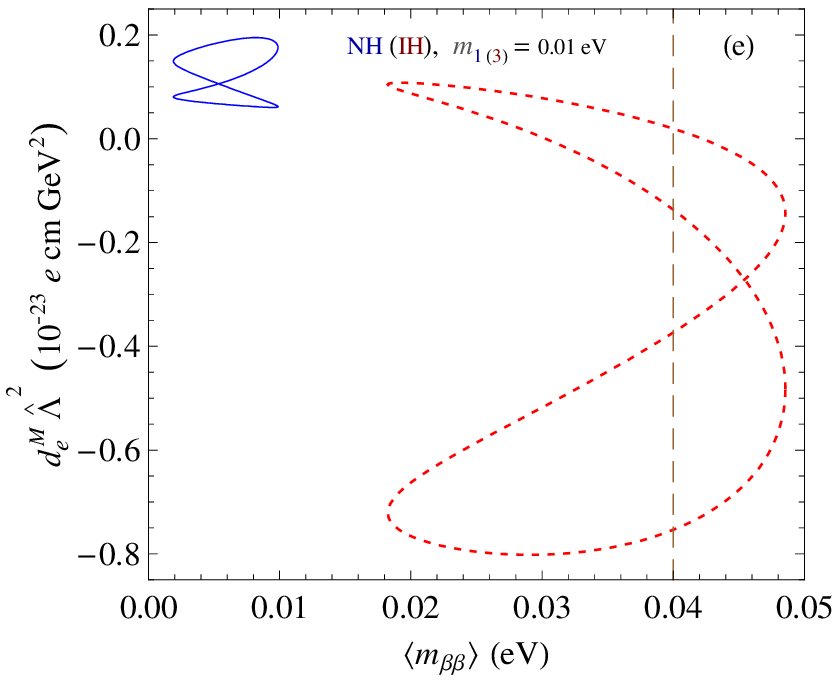} \hspace{-12pt}
\includegraphics[height=137pt,width=164pt]{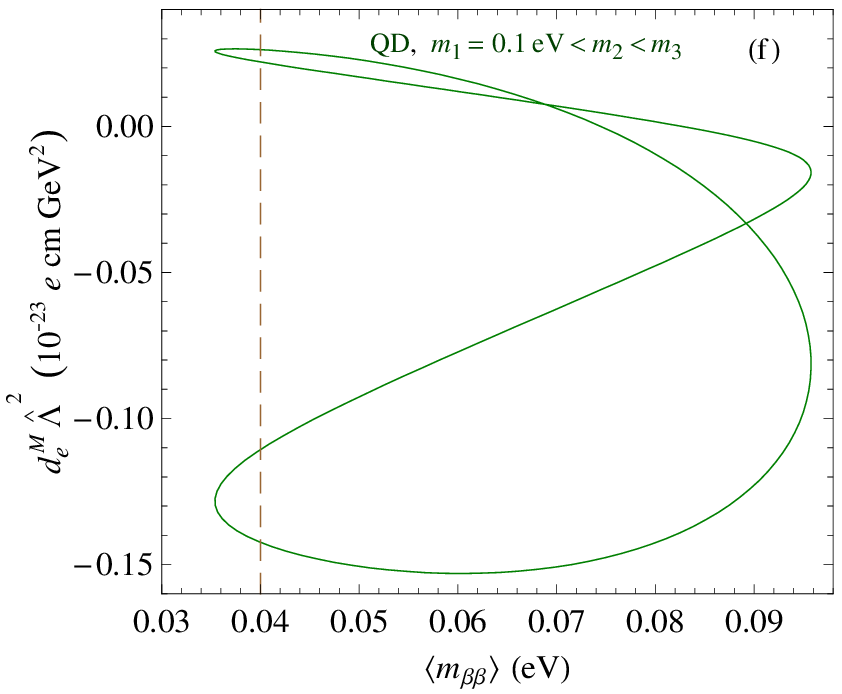}\vspace{-7pt}
\caption{Sample correlations between \,$d_e^{\rm M}\hat\Lambda^2$\, and
$\langle m_{\beta\beta}\rangle$ over \,$0\le\alpha_2^{}\le4\pi$\, for \,$\alpha_1^{}=0$\, and
the central values of $\delta$ in the cases of (a,b,c)~degenerate $\nu_{k,R}^{}$ and
\,$O=e^{i\sf R}$\, and (d,e,f)~nondegenerate $\nu_{k,R}^{}$ and \,$O=e^{\sf R}$,\,
all with $r_1^{}=2r_2^{}=3r_3^{}=\frac{1}{2}$,\, as described in the text.
The vertical dashed lines mark a possible sensitivity in future searches for
neutrinoless double-$\beta$ within decay the next decade.\label{de-mbb}}
\end{figure}

From Figs.~\ref{de-alpha-complo}-\ref{mbetabeta}, one can conclude that \,$d_e^{\rm M}$\, and
$\langle m_{\beta\beta}\rangle$ may be correlated.
For the MFV scenario under consideration and the parameter choices we made with the central
values from Table$\;$\ref{nudata}, we show in Fig.$\;$\ref{de-mbb} some sample relations
between the two observables.
One can see in particular that the plots in Fig.$\;$\ref{de-mbb}(a,c) [(d,f)] are related to
the solid curves in the first and third (green) graphs of~Fig.$\;$\ref{mbetabeta}, respectively,
and the corresponding solid curves in Fig.$\;$\ref{de-alpha-complo}
[Fig.$\;$\ref{de-alpha-realo}] for \,$r_1^{}=2r_2^{}=3r_3^{}=\frac{1}{2}$.\,
In Fig.$\;$\ref{de-mbb} we have also indicated a projected sensitivity of \,0.04\,eV\, in
future hunts for neutrinoless double-$\beta$ decay which may be achieved after several years.

The illustrations in Fig.$\;$\ref{de-mbb} suggest that, if searches in coming years still
yield null results, the acquired limits on $d_e$ and $\langle m_{\beta\beta}\rangle$ will impose
significant restrictions on various scenarios based on lepton~MFV.
On the other hand, unambiguous observations of $d_e^{\rm M}$ and/or neutrinoless double-beta
decay will help pin down the favored underlying model and parameter space, under the assumption
that the latter process is mediated by a light Majorana neutrino~\cite{nureview}.
The information to be gained from the direct neutrino-mass determination in planned tritium
$\beta$-decay experiments, with expected sensitivities as low as \,0.2\,eV~\cite{nureview},
and the total neutrino mass to be inferred from upcoming cosmological data with improved
precision will supply complementary constraints and cross checks.

Before moving on, we would like to make some remarks on the situation in which only
two right-handed neutrinos are added into the theory.
In that case, $Y_\nu$ and $M_\nu$ as defined in Eq.\,(\ref{Lm}) are 3$\times$2 and 2$\times$2
matrices, respectively.
As a natural consequence~\cite{Xing:2003ic}, it is straightforward to realize from Eq.\,(\ref{mnu})
that \,$|{\rm Det}_{\,}m_\nu^{}|=m_1^{}m_2^{}m_3^{}=0$,\, indicating that
one of $m_{1,2,3}^{}$ has to vanish.
Another difference is that the $O$ matrix in Eq.\,(\ref{Ynu}) is now 3$\times$2.
Accordingly, with \,$m_1^{}=0$\, or \,$m_3^{}=0$\, we can write respectively~\cite{He:2013cea}
\begin{eqnarray}
O \,\,=\, \left ( \begin{array}{cc} 0 & 0 \\ 1 & 0 \\ 0 & 1 \end{array} \right) O_2^{} ~~~~~
{\rm or} ~~~~~ O \,\,=\, \left ( \begin{array}{cc} 1 & 0 \\ 0 & 1 \\ 0 & 0 \end{array} \right) O_2^{}
\end{eqnarray}
where $O_2^{}$ is a complex 2$\times$2 matrix satisfying
\,$O_2^{}O_2^{\rm T}=\openone_2^{}$,\, where $\openone_2^{}$ is 2$\times$2 unit matrix.
Thus $O_2^{}$ has 2 free real parameters, whereas $O$ in the presence of 3 right-handed
neutrinos has six.
All this implies that the specific examples we have provided so far with $m_1^{}$ or $m_3^{}$
set to zero are applicable to the situation with only 2 right-handed neutrinos, as the 2 parameters
of $O_2^{}$ are functions of the 6 parameters of $O$ in the case of 3 right-handed
neutrinos with \,$m_{1\rm\,or\,3}^{}=0$.\,
We conclude that for $d_e$ the situations with 2 and 3 right-handed neutrinos are similar.

\subsection{Muon and tau EDMs\label{dmu}}

If neutrinos are of Dirac nature, the muon and tau EDMs will be tiny, like $d_e^{\rm D}$.
Therefore here, and in the rest of the section, we suppose that neutrinos are Majorana fermions.
Furthermore, for definiteness and simplicity, we consider only the scenario in which
the right-handed neutrinos are degenerate, \,$M_\nu={\cal M}\openone$,\, and the orthogonal
matrix $O$ is real.
For the neutrino parameters, we will adopt the specific values which yielded Eq.\,(\ref{deM'}) and
\,${\cal M}=6.16\;(6.22)\times 10^{14}\,$GeV\, in the NH (IH) case with \,$m_{1(3)}=0$.\,

Accordingly, from Eq.\,(\ref{dem}) we easily infer the muon and tau EDMs, respectively, to be
\begin{eqnarray} \label{dmuM}
d_\mu^{\rm M} &\,=\,& d_e^{\rm M}\; \frac{m_\mu^{}\bigl(m_\tau^2-m_e^2\bigr)}
{m_e^{}\bigl(m_\mu^2-m_\tau^2\bigr)} ~, ~~~~~~~
d_\tau^{\rm M} \,\,=\,\, d_e^{\rm M}\; \frac{m_\tau^{}\bigl(m_e^2-m_\mu^2\bigr)}
{m_e^{}\bigl(m_\mu^2-m_\tau^2\bigr)} ~,
\end{eqnarray}
with $d_e^{\rm M}$ in Eq.\,(\ref{deM'}).
Since \,$d_\tau^{\rm M}\sim-0.06\,d_\mu^{\rm M}$\, and the experimental information
on $d_\tau$ is still imprecise~\cite{pdg}, we will not deal with it further.

Hence we get~\,$d_\mu^{\rm M}=-2.3\;(-0.26)\times10^{-21}{\rm\;GeV}^2/\hat\Lambda^2$.\,
Currently the best measured limit on the muon EDM is
\,$|d_\mu|_{\rm exp}<1.8\times10^{-19\;}e$\,cm\, at 95\% CL, set by the Muon $(g-2)$
Collaboration~\cite{Bennett:2008dy}.
This implies
\begin{eqnarray} \label{lambdamu}
\hat\Lambda \,\,>\,\, 0.11\;(0.038){\rm\;GeV} ~,
\end{eqnarray}
which are not competitive to the bounds in Eq.\,(\ref{hatLambda0}) from $|d_e|_{\rm exp}$.

\subsection{$\bm{CP}$-violating electron-neutron interactions\label{eenn}}

Searches for atomic and molecular EDMs may be sensitive to other mechanisms
possibly responsible for them besides the electron EDM, such as the EDMs of nuclei
and $CP$-violating electron-nucleon interactions.
In this section we are interested in the third possibility, particularly that
described by~\cite{Engel:2013lsa,Ginges:2003qt}
\begin{eqnarray}
{\cal L}_{eN} \,\,=\,\, \frac{-i C_S G_{\rm F}}{\sqrt2}\,\bar e\gamma_5^{}e\,\bar N N \,-\,
\frac{iC_P G_{\rm F}}{\sqrt2}\,\bar e e\,\bar N\gamma_5^{}N \,-\,
\frac{i C_T G_{\rm F}}{\sqrt2}\,\bar e\sigma^{\kappa\omega}\gamma_5^{}e\,
\bar N\sigma_{\kappa\omega}^{}N ~.
\end{eqnarray}
The recent ACME experiment has set the best limit on the first coupling,
\,$|C_S|_{\rm exp}<5.9\times10^{-9}$\, at 90\%\,\,CL~\cite{acme}.
The strictest limits on the other two, \,$|C_P|_{\rm exp}<5.1\times10^{-7}$\, and
\,$|C_T|_{\rm exp}<1.5\times10^{-9}$\, at 95\%\,\,CL, were based on the latest
search for the EDM of the $^{199}$Hg atom~\cite{Griffith:2009zz}.

These interactions may originate from MFV in the lepton sector as well as the quark sector,
which has to be included for a~consistent analysis.
The Lagrangian for the relevant lowest-order operators is
\begin{eqnarray}
{\cal L}_{\ell q} &\,=\,&
\frac{1}{\Lambda^2} \Bigl( \bar U_R^{}Y_u^\dagger\bar\Delta_{qu1}^{}i\tau_2^{}Q_L^{}\,
\bar E_R^{}Y_e^\dagger\bar\Delta_{\ell1}^{}L_L^{}
+ \bar Q_L^{}\bar\Delta_{qd1}^\dagger Y_{d\,}^{}D_R^{}\,
\bar E_R^{}Y_e^\dagger\bar\Delta_{\ell2\,}^{}L_L^{}
\nonumber \\ && ~~~~ \,+\,
\bar U_R^{}\sigma^{\kappa\omega}Y_u^\dagger\bar\Delta_{qu2\,}^{}i\tau_2^{}Q_L^{}\,
\bar E_R^{}\sigma_{\kappa\omega}^{}Y_e^\dagger\bar\Delta_{\ell3}^{}L_L^{}
\nonumber \\ && ~~~~ \,+\,
\bar Q_L^{}\sigma^{\kappa\omega}\bar\Delta_{qd2}^\dagger Y_{d\,}^{}D_R^{}\,
\bar E_R^{}\sigma_{\kappa\omega}^{}Y_e^\dagger\bar\Delta_{\ell4\,}^{}L_L^{} \Bigr)
\,+\; {\rm H.c.} ~,
\end{eqnarray}
where $\bar\Delta_{qu\varsigma,qd\varsigma}$ $\bigl(\bar\Delta_{\ell1,\ell2,\ell3,\ell4}\bigr)$
are the same in form as $\Delta$ in Eq.\,(\ref{general}) and contain the quark (lepton) Yukawa
couplings.
The leptonic contributions to $C_{S,P,T}$ turn out to be dominant.

To determine $C_S$, we need the matrix elements
\,$\langle N|m_q^{}\bar q q|N\rangle=g_q^N\bar u_N^{}u_N^{}v$.\,
Thus, we derive
\begin{eqnarray} \label{cs}
C_S^{} &\,=\,& \frac{16\sqrt2\,m_{e\,}^{}{\cal M}^3}{\Lambda^{2\,}G_{\rm F\,}^{}v^9}
\bigl(m_\tau^2-m_\mu^2\bigr)
\bigl(m_1^{}-m_2^{}\bigr)\bigl(m_2^{}-m_3^{}\bigr)\bigl(m_3^{}-m_1^{}\bigr)
\nonumber \\ && \times~
\Bigl[ \bigl(g_u^N+g_c^N+\kappa_{u1\,}^{}g_t^N\bigr)\,\bar\xi_{12}^{\ell1}
- \bigl(g_d^N+g_s^N+\kappa_{d1\,}^{}g_b^N\bigr)\,\bar\xi_{12}^{\ell2} \Bigr] J_\ell^{} ~,
\end{eqnarray}
where $\bar\xi_{12}^{\ell1,\ell2}$ belong to $\bar\Delta_{\ell1,\ell2}$ and have absorbed
the first coefficients $\bar\xi_1^{u1,d1}$ of $\bar\Delta_{qu1,qd1}$, respectively,
and \,$\kappa_x^{}\simeq1+\bigl(\bar\xi_2^x+\bar\xi_4^x\bigr)/\bar\xi_1^x$\, are
numbers expected to be at most of~${\cal O}(1)$.
Numerically, we adopt the chiral Lagrangian estimate~\cite{He:2008qm}
\begin{eqnarray}
g_u^N &=\,\, 0.04~(0.12)\times10^{-3} ~, ~~~~~~~~~ g_d^N &=\,\, 0.08~(0.21)\times10^{-3} ~, \\
g_s^N &=\,\, 0.25~(2.88)\times10^{-3} ~, ~~~~~~~ g_{c,b,t}^N &=\,\, 0.26~(0.05)\times10^{-3} ~,
\end{eqnarray}
corresponding to the so-called pion-nucleon sigma term \,$\sigma_{\pi N}^{}=30~(80)~$MeV,\,
which is not yet well-determined~\cite{lattice,Alarcon:2012nr}.\footnote{\baselineskip=12pt
Lattice QCD computations~\cite{lattice} tend to produce results smaller than those of chiral
Lagrangian calculations and some other methods~\cite{Alarcon:2012nr}.
As a consequence, employing the lattice values of $g_q^N$ in Eq.\,(\ref{cs}) would yield even
looser limits than in~Eq.\,(\ref{hatLambda}).}
Then, using the maxima of $g_q^N$ and assuming \,$\kappa_x^{}=1$,\,
we can neglect the $\bar\xi_{12}^{\ell1}$ part in Eq.\,(\ref{cs}) to obtain from
\,$|C_S|_{\rm exp}<5.9\times10^{-9}$\,
\begin{eqnarray} \label{hatLambda}
\frac{\Lambda}{\bigl|\bar\xi_{12}^{\ell2}\bigr|\raisebox{1pt}{$^{1/2}$}} \,\,>\,\,
0.27\;(0.091){\rm\;GeV}
\end{eqnarray}
in the NH (IH) neutrino parameter values specified in the preceding subsection.
These restraints are far weaker than those from $|d_e|_{\rm exp}$.

For $C_P$, the expression is the same as that for $C_S$ in Eq.\,(\ref{cs}), except
\,$g_q^N$\, is replaced by \,$\varsigma_{q\,}^{}h_q^N m_N^{}/v$\, with
\,$\varsigma_q^{}=+1\,(-1)$\, if \,$q=u,c,t\,(d,s,b)$\, and $h_q^N$ defined by
\,$\langle N|m_q^{}\bar q\gamma_5^{}q|N\rangle=h_q^N m_{N\,}^{}\bar u_N^{}\gamma_5^{}u_N^{}$.\,
Since for mercury $C_P$ is estimated to be mostly from the neutron
contribution~\cite{Ginges:2003qt}, we focus on it.
Ignoring the effects of~$h_{c,b,t}^n$, we can relate $h_{u,d,s}^n$ to the axial-vector charges
$g_A^{(0,3,8)}$ by $6h_u^n=2g_A^{(0)}-3g_A^{(3)}+g_A^{(8)}$,\, $6h_d^n=2g_A^{(0)}+3g_A^{(3)}+g_A^{(8)}$,\, and
\,$3h_s^n=g_A^{(0)}-g_A^{(8)}$,\, where \,$g_A^{(0)}=0.33\pm0.06$,\,
$g_A^{(3)}=1.270\pm0.003$,\, and \,$g_A^{(8)}=0.58\pm0.03$\, were measured in baryon
$\beta$-decay and deep inelastic scattering experiments~\cite{Aidala:2012mv}.
Taking \,$\bar\xi_{12}^{\ell1}=\bar\xi_{12}^{\ell2}$\, and maximizing $C_P$, we obtain
from \,$|C_P|_{\rm exp}<5.1\times10^{-7}$\,
\begin{eqnarray} \label{Lambdahat}
\frac{\Lambda}{\bigl|\bar\xi_{12}^{\ell2}\bigr|\raisebox{1pt}{$^{1/2}$}} \,\,>\,\,
0.020\;(0.0068){\rm\;GeV} \,,
\end{eqnarray}
less restrictive than Eq.\,(\ref{hatLambda}) by more than an order of magnitude.

To evaluate $C_T$, we need the matrix elements
\,$\langle N|\bar q\sigma^{\kappa\omega}q|N\rangle=
\rho_{N\,}^q\bar u_N^{}\sigma^{\kappa\omega}u_N^{}$,\,
where $\rho_N^q$ have the values in Eq.\,(\ref{rhonq}) for light quarks, assuming isospin
symmetry, and vanish for heavier quarks.
This leads us to
\begin{eqnarray}
C_T^{} &\,=\,& \frac{32\sqrt2\,m_{e\,}^{}{\cal M}^3}{\Lambda^{2\,}G_{\rm F\,}^{}v^{10}}
\bigl(m_\tau^2-m_\mu^2\bigr)
\bigl(m_1^{}-m_2^{}\bigr)\bigl(m_2^{}-m_3^{}\bigr)\bigl(m_3^{}-m_1^{}\bigr)\,
\rho_{u\,}^n m_u^{}\,\bar\xi_{12}^{\ell3}\, J_\ell^{} ~,
\end{eqnarray}
where $\bar\xi_{12}^{\ell3}$ belongs to $\bar\Delta_{\ell3}$ and has absorbed
$\bar\xi_1^{u2}$ from $\bar\Delta_{qu2}$.
The contributions of the down-type quarks cancel due to the relation
\,$\bar q\sigma^{\kappa\omega}\gamma_5^{}q\,\bar e\sigma_{\kappa\omega}e=
\bar q\sigma^{\kappa\omega}q\,\bar e\sigma_{\kappa\omega}\gamma_5^{}e$.\,
Hence, with the largest $m_u^{}$ from Section\,\,\ref{nedm} and \,$\rho_n^u=-0.78$,\,
we get from \,$|C_T|_{\rm exp}<1.5\times10^{-9}$\,
\begin{eqnarray}
\frac{\Lambda}{\bigl|\bar\xi_{12}^{\ell4}\bigr|\raisebox{1pt}{$^{1/2}$}} \,\,>\,\,
0.033\;(0.011){\rm\;GeV} \,,
\end{eqnarray}
comparable to Eq.\,(\ref{Lambdahat}).

\subsection{Muon $\bm{g-2}$, $\bm{\mu\to e\gamma}$,
nuclear $\bm{\mu\to e}$ conversion, $\bm{\bar{B}\to X_s\gamma}$\label{ll'g}}

The MFV coefficient $\xi_{12}^\ell$ that determines the electron EDM also enters
the anomalous magnetic moment of the muon $(g_\mu-2)$ and the rates of the radiative
decay \,$\mu \to e\gamma$\, and nuclear \,$\mu\to e$\, conversion,
the latter two being still unobserved.
Since $g_\mu-2$ has been very precisely measured and the experimental limits of
the flavor-changing transitions are stringent, it is important to check if these processes can
yield stronger bounds on \,$\hat\Lambda=\Lambda/\bigl|\xi_{12}^\ell\bigr|\raisebox{1pt}{$^{1/2}$}$\,
than those evaluated in the preceding subsections.
Although the other $\xi_{r\neq12,16}^\ell$ terms may contribute to these processes as well and
therefore may reduce the impact of the $\xi_{12}^\ell$ term, one also cannot rule out
the possibility of a~scenario in which the latter dominates the other contributions.

The anomalous magnetic moment $a_l^{}$ of lepton $l$ is described by
\,${\cal L}_{a_l}=\bigl[e\,a_l/(4m_l)]\,\bar l\sigma^{\kappa\omega}l F_{\kappa\omega}$.\,
From Eq.\,(\ref{Leff}) we have
\begin{eqnarray} \label{Ll2l'g}
{\cal L}_{E_i\to E_k\gamma} \,\,=\,\, \frac{e}{2\Lambda^2}\, \bar E_{k\,}^{}
\sigma_{\kappa\omega}^{} \Bigl\{
m_{E_k}^{}(\Delta_\ell)_{ki}^{} + m_{E_i}^{}(\Delta_\ell)_{ik}^* -
\Bigl[ m_{E_k}^{}(\Delta_\ell)_{ki}^{}-m_{E_i}^{}(\Delta_\ell)_{ik}^*\Bigr]\gamma_5^{}
\Bigr\} E_{i\,}^{}F^{\kappa\omega} ~,
\end{eqnarray}
where \,$(E_1,E_2,E_3)=(e,\mu,\tau)$\, and \,$\Delta_\ell=\Delta_{\ell1}-\Delta_{\ell2}$.\,
It follows that
\begin{eqnarray} \label{al}
a_{E_k}^{} \,\,=\,\, \frac{4 m_{E_k}^2}{\Lambda^2}\, {\rm Re}(\Delta_\ell)_{kk}^{} ~.
\end{eqnarray}
Thus, with the NH neutrino parameter values specified in Section\,\,\ref{dmu}, we have
\begin{eqnarray} \label{amu}
a_\mu^{} \,=\, \frac{4 m_\mu^2}{\Lambda^2}\, {\rm Re}(\Delta_\ell)_{22} \,=\,
\Bigl( 45_{\,}\xi_1^\ell+23_{\,}\xi_2^\ell+20_{\,}\xi_4^\ell+0.00085_{\,}\xi_8^\ell
+ 0.00094_{\,}\xi_{12}^\ell \Bigr) \frac{\rm GeV^2}{10^3\Lambda^2}
\end{eqnarray}
where terms with numerical factors much smaller than that of $\xi_{12}^\ell$ have been dropped.
The corresponding numbers in the IH case are roughly similar.
Currently the experimental and SM values differ by
\,$a_\mu^{\rm exp}-a_\mu^{\scriptscriptstyle\rm SM}=(249\pm87)\times10^{-11}$\,~\cite{Aoyama:2012wk},
which suggests that we can require \,$\bigl|a_\mu^{}\bigr|<3.4\times10^{-9}$.\,
For the $\xi_{12}^\ell$ term alone, this translates into the rather loose limit
\,$\hat\Lambda>17\;\rm GeV$,\, which may be weakened in the presence of the other terms
in Eq.\,(\ref{amu}).

From Eq.\,(\ref{Ll2l'g}), one can also calculate the branching ratio ${\cal B}(\mu\to e\gamma)$
of \,$\mu\to e\gamma$.\,
In the \,$m_e^{}=0$\, limit
\begin{eqnarray} \label{m2eg} \displaystyle
{\cal B}(\mu\to e\gamma) \,\,=\,\, \frac{\tau_{\mu\,}^{}e_{\;\;}^2 m_\mu^5}{4\pi_{\,}\Lambda^4}
\bigl|(\Delta_\ell)_{21}^{}\bigr|\raisebox{1pt}{$^2$} ~,
\end{eqnarray}
where $\tau_\mu^{}$ is the muon lifetime.
In the NH case
\begin{eqnarray} \label{Deltal21}
(\Delta_\ell)_{21} \,=\, (0.061-0.1i)\xi_2^\ell+(0.011-0.11i)\xi_4^\ell
- \bigl[ (26+48i)\xi_8^\ell + (6+51i)\xi_{12}^\ell \bigr]\!\times\!10^{-7} ~,
\end{eqnarray}
where again terms with numerical factors less than that of $\xi_{12}^\ell$ have been ignored.
The $(\Delta_\ell)_{21}$ numbers in the IH case are comparable in size.
If only $\xi_{12}^\ell$ is nonvanishing in $(\Delta_\ell)_{21}$, then the experimental bound
\,${\cal B}(\mu\to e\gamma)_{\rm exp}<5.7\times10^{-13}$\,~\cite{meg} implies
\begin{eqnarray} \label{hatLambda'}
\hat\Lambda \,\,>\,\, 2.0\;\rm TeV ~.
\end{eqnarray}
This is stronger by up to \mbox{\footnotesize\,$\sim$\,}20 times than those in
Eq.\,(\ref{hatLambda0}) from the electron EDM data.
However, the other terms in $(\Delta_\ell)_{21}$, some of which are potentially
much bigger than the $\xi_{12}^\ell$ contribution, can in principle decrease the impact of the latter,
thereby lessening the restriction on $\hat\Lambda$.
Consequently, $d_e$ provides a less ambiguous probe for $\hat\Lambda$.

Measurements on \,$\mu\to e$\, conversion in nuclei can provide constraints on new physics
competitive to those from \,$\mu\to e\gamma$\, searches~\cite{deGouvea:2013zba}.
The relation between the rates of \,$\mu\to e$\, conversion and \,$\mu\to e\gamma$\,
produced by possible new physics is available from Ref.\,\cite{Kitano:2002mt}.
Assuming that the MFV dipole interactions described by Eq.\,(\ref{operators'}) saturate
\,$\mu\to e$\, conversion in nucleus $\cal N$, we can express its rate divided by the rate
$\omega_{\rm capt}^{\cal N}$ of $\mu$~capture in $\cal N$ as
\begin{eqnarray} \label{muN2eN}
{\cal B}(\mu{\cal N}\to e{\cal N}) \,\,=\,\,
\frac{e^2 m_{\mu\,}^5\bigl|(\Delta_\ell)_{21\,}^{}D_{\cal N}^{}\bigr|^2}
{4\Lambda^{4\,}\omega_{\rm capt}^{\cal N}} ~,
\end{eqnarray}
where $D_{\cal N}$ represents the dimensionless overlap integral for~$\cal N$ and
for the NH parameter choices $(\Delta_\ell)_{21}$ is given in~Eq.\,(\ref{Deltal21}).
Based on the existing experimental limits on \,$\mu\to e$\, transition in various nuclei~\cite{pdg}
and the corresponding $D_{\cal N}$ and $\omega_{\rm capt}^{\cal N}$ values~\cite{Kitano:2002mt},
significant restrictions can be expected from
\,${\cal B}(\mu_{\,}{\rm Ti}\to e_{\,}{\rm Ti})_{\rm exp}<6.1\times10^{-13}$~\cite{Papoulias:2013gha}
and \,${\cal B}(\mu_{\,}{\rm Au}\to e_{\,}{\rm Au})_{\rm exp}<7\times10^{-13}$~\cite{pdg}.
From these data, if only the $\xi_{12}^\ell$ term in $(\Delta_\ell)_{21}$ is nonvanishing,
employing \,$D_{\rm Ti}=0.087$,\, $D_{\rm Au}=0.189$, \,$\omega_{\rm capt}^{\rm Ti}=2.59\times10^6/\rm s$,\,
and \,$\omega_{\rm capt}^{\rm Au}=13.07\times10^6/\rm s$\,~\cite{Kitano:2002mt}, we extract
\begin{eqnarray} \label{hatLambda''}
\hat\Lambda_{\rm Ti}^{} \,\,>\,\, 0.49\;\rm TeV ~, ~~~~~~~
\hat\Lambda_{\rm Au}^{} \,\,>\,\, 0.47\;\rm TeV ~,
\end{eqnarray}
which are stricter than the results in Eq.\,(\ref{hatLambda}) by up to a few times,
but weaker than~Eq.\,(\ref{hatLambda'}).
Upcoming searches for \,$\mu\to e$\, in the next several years will, if it still eludes
detection, lower the limits to the $10^{-16}$ level or better~\cite{deGouvea:2013zba},
which will push $\hat\Lambda$ higher.
Nevertheless, since again the other $\xi_r^\ell$ terms are generally present in
$(\Delta_\ell)_{21}$, these bounds on $\hat\Lambda$ are not unambiguous.
Thus $d_e$ provides the best probe for $\hat\Lambda$ in connection with $CP$ violation.

Since there is a possibility that the MFV scales in the lepton and quark sectors are equal
or related to each other, it is of interest to check if there are any quark processes
that can also offer bounds stronger than those on $\hat\Lambda$ from~$d_e$.
Since, as we saw in Section\,\,\ref{nedm}, the neutron EDM could not provide a competitive
constraint, we need to look at other processes.
The most stringent restriction on the quark MFV scale turns out to be from the rare decay
\,$\bar B\to X_s\gamma$\,~\cite{D'Ambrosio:2002ex}.
Its experimental and SM branching ratios are
\,${\cal B}\bigl(\bar B\to X_s\gamma\bigr){}_{\rm exp}=(3.43\pm0.22)\times10^{-4}$\,~\cite{hfag}
and \,${\cal B}\bigl(\bar B\to X_s\gamma\bigr){}_{\scriptscriptstyle\rm SM}=
(3.15\pm0.23)\times10^{-4}$\,~\cite{Misiak:2006ab}
both for the photon energy \,$E_\gamma>1.6$\,GeV.\,
To isolate the MFV contribution, we adopt from Ref.\,\cite{D'Ambrosio:2002ex} the relation
\begin{eqnarray} \label{b2sg} \displaystyle
{\cal B}\bigl(\bar B\to X_s\gamma\bigr){}_{\rm exp} \,\,\simeq\,\,
\bigl(1-2.4_{\,}C_{7\gamma}^{\scriptscriptstyle\rm MFV}\bigr)
{\cal B}\bigl(\bar B\to X_s\gamma\bigr){}_{\scriptscriptstyle\rm SM} ~,
\end{eqnarray}
where $C_{7\gamma}^{\scriptscriptstyle\rm MFV}$ is evaluated at \,$\mu=m_W^{}$\, and
enters the effective Lagrangian
\begin{eqnarray} \label{Lb2sg}
{\cal L}_{b\to s\gamma}^{} \,\,=\,\, \frac{e G_{\rm F}^{\;~}m_b^{}}{8\sqrt2\,\pi^2}\,V_{ts}^*V_{tb\,}^{}
\bigl(C_{7\gamma}^{\scriptscriptstyle\rm SM}+C_{7\gamma}^{\scriptscriptstyle\rm MFV}\bigr)\,
\bar s\sigma^{\kappa\omega}\bigl(1+\gamma_5^{}\bigr)b\,F_{\kappa\omega} ~,
\end{eqnarray}
implying that
\begin{eqnarray}
C_{7\gamma}^{\scriptscriptstyle\rm MFV} \,\,=\,\, \frac{4\sqrt2\,\pi^2}{\Lambda^{2\,}G_{\rm F}^{}}
\frac{(\Delta_{qd})_{32}^*}{V_{ts}^*V_{tb}^{}} ~.
\end{eqnarray}
For the central values of the quark masses quoted in Section\,\,\ref{nedm}
\begin{eqnarray} \label{Deltaqd}
\frac{(\Delta_{qd})_{32}^*}{V_{ts}^*V_{tb}^{}} &\,=\,& \xi_{2\,}^d y_t^2+\xi_{4\,}^d y_t^4
+ y_{b\,}^2 \bigl( \xi_{7\,}^d y_t^2 + \xi_{8\,}^d y_t^4 + \xi_{9\,}^d y_t^4
+ \xi_{12\,}^d y_t^6\bigr) ~,
\end{eqnarray}
where the imaginary parts and other $\xi_r^d$ terms are negligible, \,$y_t^2\simeq1$,\,
and \,$y_b^2\simeq0.0003$.\,
Combining the errors in quadrature for the ratio of branching ratios in Eq.\,(\ref{b2sg}) and
assuming that \,$\xi_{r\neq12}^d=0$,\, we obtain at~90\%\,CL
\begin{eqnarray}
\frac{\Lambda}{\bigl|\xi_{12}^{d}\bigr|\raisebox{1pt}{$^{1/2}$}} \,\,>\,\, 0.19\;(0.11){\rm\;TeV}
\end{eqnarray}
if $C_{7\gamma}^{\scriptscriptstyle\rm MFV}$ has destructive (constructive) interference with
the SM term.
These numbers are somewhat lower than those in Eq.\,(\ref{hatLambda0}) and, as in the lepton cases,
may go down in the presence of the other $\xi_r^d$ terms in Eq.\,(\ref{Deltaqd}).

\section{Conclusions\label{conclusion}}

We have explored $CP$ violation beyond the SM via fermion EDMs under
the framework of minimal flavor violation.
The new physics scenarios covered are the standard model slightly expanded with
the addition of three right-handed neutrinos and its extension including the seesaw mechanism
for endowing neutrinos with light mass.
Addressing the quark sector first, we find that the present empirical limit on the neutron EDM
implies only a loose constraint on the scale of quark~MFV.
Moreover, we show that the impact of MFV on the contribution of the strong theta-term to
the neutron EDM is insignificant.
Turning to the lepton sector, we demonstrate that the current EDM data also yield
unimportant restraints on the leptonic MFV scale if neutrinos are of Dirac nature.
In contrast, if neutrinos are Majorana particles, the constraints become tremendously more
stringent and, in light of the latest search for $d_e$ by ACME, restrict the MFV scale to
above a~few hundred GeV or more.
Furthermore, $d_e$ can be connected in a complementary way to neutrinoless double-$\beta$
decay if it is induced mainly or solely by the exchange of a light Majorana neutrino.
We find in addition that constraints on the MFV scale inferred from the $CP$-violating
electron-nucleon couplings probed by ACME and the most recent search for the EDM of mercury
are relatively weak as well.
Finally, we take into account potential restrictions from the measurements on the muon~$g-2$,
radiative decays \,$\mu\to e\gamma$\, and \,$\bar B\to X_s\gamma$,\, and
\,$\mu\to e$\, conversion in nuclei, which are not sensitive to $CP$ violation.

\acknowledgments

This research was supported in part by the MOE Academic Excellence Program (Grant No. 102R891505)
and NSC of ROC and by NNSF (Grant No. 11175115) and Shanghai Science and Technology Commission
(Grant No. 11DZ2260700) of PRC.

\appendix

\section{Evaluation of some products of $\bm{\textsf A}$ and $\bm{\textsf B}$ matrices\label{ABproducts}}

From the Cayley-Hamilton identity in Eq.\,(\ref{ch}) with \,$X=a{\sf A}+b{\sf B}$,\, where
and $a$ and $b$ are free parameters, one can extract~\cite{Colangelo:2008qp}
\begin{eqnarray} \label{aab}
{\sf A}^2{\sf B}+{\sf ABA}+{\sf BA}^2 &\,=\,&
{\sf A}^2\langle{\sf B}\rangle+({\sf AB}+{\sf BA})\langle{\sf A}\rangle+
{\sf A}(\langle{\sf AB}\rangle-\langle{\sf A}\rangle\langle{\sf B}\rangle)+
\mbox{$\frac{1}{2}$}\bigl(\bigl\langle{\sf A}^2\bigr\rangle-\langle{\sf A}\rangle^2\bigr){\sf B}
\nonumber \\ && \!+~ \openone \Bigl[
\mbox{$\frac{1}{2}$}\bigl(\langle{\sf A}\rangle^2-\bigl\langle{\sf A}^2\bigr\rangle\bigr)
\langle{\sf B}\rangle
+ \bigl\langle{\sf A}^2{\sf B}\bigr\rangle - \langle{\sf A}\rangle\langle{\sf AB}\rangle \Bigr]
\end{eqnarray}
and an analogous expression for \,${\sf ABB}+{\sf BAB}+{\sf BBA}$,\, where
\,$\langle\cdots\rangle={\rm Tr}(\cdots)$.\,
These relations can be used to derive other combinations of $\sf A$ and $\sf B$.
For instance, by replacing $\sf B$ with ${\sf B}^2$ ($[{\sf B},{\sf AB}]$)
in~Eq.\,(\ref{aab}), we can write
\,${\sf A}^2{\sf B}^2+{\sf A}{\sf B}^2{\sf A}+{\sf B}^2{\sf A}^2$\,
$\bigl({\sf A}^2{\sf BAB}+{\sf ABABA}+{\sf BABA}^2
-{\sf A}^2{\sf B}^2{\sf A}-{\sf A}{\sf B}^2{\sf A}^2-{\sf A}^3{\sf B}^2\bigr)$
in terms of lower-ordered products of these matrices.
After further algebra, we arrive at
\begin{eqnarray}
{\sf A}^2{\sf BAB}^2 &\,=\,& \zeta_1^{}\openone + \zeta_{2\,}^{}{\sf A} + \zeta_{3\,}^{}{\sf B} +
\zeta_{4\,}^{}{\sf A}^2 + \zeta_{5\,}^{}{\sf B}^2 + \zeta_{6\,}^{}{\sf A B} +
\zeta_{7\,}^{}{\sf B A} + \zeta_{8\,}^{}{\sf ABA} + \zeta_{9\,}^{}{\sf B A}^2
\nonumber \\ && \! +~
\zeta_{10\,}^{}{\sf BAB} + \zeta_{11\,}^{}{\sf AB}^2 +
\zeta_{12\,}^{}{\sf ABA}^2 + \zeta_{13\,}^{}{\sf A}^2{\sf B}^2 +
\zeta_{14\,}^{}{\sf B}^2{\sf A}^2 + \zeta_{15\,}^{}{\sf B}^2{\sf AB}
\nonumber \\ && \! +~
\zeta_{16\,}^{}{\sf AB}^2{\sf A}^2 + \zeta_{17\,}^{}{\sf B}^2{\sf A}^2{\sf B} ~, ~~~~~~~
\end{eqnarray}
where
\begin{eqnarray}
\zeta_1^{} &=& \frac{\bigl\langle{\sf A}^2{\sf BAB}^2\bigr\rangle +
\langle{\sf AB}\rangle \bigl\langle{\sf A}^2{\sf B}^2\bigr\rangle +
\bigl\langle{\sf A}^2{\sf B}\bigr\rangle \bigl\langle{\sf AB}^2\bigr\rangle}{3}
+ \langle{\sf A}\rangle\langle{\sf B}\rangle
\frac{4\bigl\langle{\sf A}^2\bigr\rangle \bigl\langle{\sf B}^2\bigr\rangle -
6\bigl\langle{\sf A}^2{\sf B}^2\bigr\rangle -
3\langle{\sf A}\rangle^2\langle\sf B\rangle^2}{6}
\nonumber \\ && \! +~
\frac{\bigl(\langle{\sf B}\rangle^3-\langle{\sf B}\rangle \bigl\langle{\sf B}^2\bigr\rangle\bigr)
{\rm Det}{\sf A} + \bigl(\langle{\sf A}\rangle^3
- \langle{\sf A}\rangle \bigl\langle{\sf A}^2\bigr\rangle\bigr){\rm Det}{\sf B}}{6}
+ \langle{\sf AB}\rangle \frac{13 \langle{\sf A}\rangle^2\langle{\sf B}\rangle^2
- 3\bigl\langle{\sf A}^2\bigr\rangle \bigl\langle{\sf B}^2\bigr\rangle}{12}
\nonumber \\ && \! +~
\langle{\sf AB}\rangle
\frac{5\langle{\sf A}\rangle^2\bigl\langle{\sf B}^2\bigr\rangle +
\langle{\sf B}\rangle^2 \bigl\langle{\sf A}^2\bigr\rangle -
8\langle{\sf A}\rangle \bigl\langle{\sf A}{\sf B}^2\bigr\rangle -
4\langle{\sf B}\rangle \bigl\langle{\sf A}^2{\sf B}\bigr\rangle}{12} +
\bigl\langle{\sf AB}^2\bigr\rangle \langle{\sf B}\rangle
\frac{\langle{\sf A}\rangle^2-\bigl\langle{\sf A}^2\bigr\rangle}{6}
\nonumber \\ && \! -~
\langle{\sf A}\rangle \bigl\langle{\sf B}^2\bigr\rangle
\frac{\langle{\sf A}\rangle^2\langle{\sf B}\rangle +
2\bigl\langle{\sf A}^2{\sf B}\bigr\rangle}{6} ~,
\end{eqnarray}
\allowdisplaybreaks
\begin{eqnarray}
\zeta_2^{} &=&
\frac{-\langle{\sf A}^2\rangle\,{\rm Det}{\sf B}}{3} +
\langle{\sf AB}\rangle \frac{4 \bigl\langle{\sf AB}^2\bigr\rangle -
5 \langle{\sf A}\rangle \bigl\langle{\sf B}\bigr\rangle^2 -
3 \langle{\sf A}\rangle \bigl\langle{\sf B}^2\bigr\rangle}{6} +
\langle{\sf B}\rangle \bigl\langle{\sf B}^2\bigr\rangle
\frac{7\langle{\sf A}\rangle^2+\bigl\langle{\sf A}^2\bigr\rangle}{12}
\nonumber \\ && \! -~
\langle{\sf B}\rangle
\frac{2\langle{\sf A}\rangle \bigl\langle{\sf AB}^2\bigr\rangle +
\bigl\langle{\sf A}^2{\sf B}^2\bigr\rangle}{3} +
\frac{\langle{\sf B}\rangle^2 \bigl\langle{\sf A}^2{\sf B}\bigr\rangle}{3} +
\langle{\sf B}\rangle^3\,
\frac{9\langle{\sf A}\rangle^2-\bigl\langle{\sf A}^2\bigr\rangle}{12} ~,
\end{eqnarray}
\begin{eqnarray}
\zeta_3^{} &\,=\,&
\frac{-\langle{\sf B}^2\rangle\,{\rm Det}{\sf A}}{3} +
\langle{\sf AB}\rangle \frac{2 \langle{\sf A}^2{\sf B}\rangle -
5 \langle{\sf A}\rangle^2 \langle{\sf B}\rangle -
\langle{\sf A}^2\bigr\rangle \langle{\sf B}\rangle}{6} +
\langle{\sf A}\rangle \bigl\langle\raisebox{-1pt}{${\sf A}^2$}\bigr\rangle
\frac{3\langle{\sf B}\rangle^2-\bigl\langle\raisebox{-1pt}{${\sf B}^2$}\bigr\rangle}{12}
\nonumber \\ && \! -~
\langle{\sf A}\rangle
\frac{\bigl\langle\raisebox{-1pt}{${\sf A}^2\sf B$}\bigr\rangle \langle{\sf B}\rangle +
\bigl\langle\raisebox{-1pt}{${\sf A}^2{\sf B}^2$}\bigr\rangle}{3} +
\bigl\langle\raisebox{-1pt}{${\sf AB}^2$}\bigr\rangle \frac{\langle{\sf A}\rangle^2 +
\bigl\langle\raisebox{-1pt}{${\sf A}^2$}\bigr\rangle}{6} +
\langle{\sf A}\rangle^3\,
\frac{9\langle{\sf B}\rangle^2+\bigl\langle\raisebox{-1pt}{${\sf B}^2$}\bigr\rangle}{12} ~,
\\ \nonumber \\
\zeta_4^{} &\,=\,& \frac{\langle{\sf A}\rangle\,{\rm Det}{\sf B}}{3} +
\bigl\langle\raisebox{-1pt}{${\sf B}^2$}\bigr\rangle
\frac{2\langle{\sf AB}\rangle-7\langle{\sf A}\rangle \langle{\sf B}\rangle}{6} -
\frac{\langle{\sf A}\rangle \langle{\sf B}\rangle^3}{6} +
\frac{\bigl\langle\raisebox{-1pt}{${\sf AB}^2$}\bigr\rangle \langle{\sf B}\rangle}{3} ~,
\\ \nonumber \\
\zeta_5^{} &\,=\,& \frac{\langle{\sf B}\rangle\,{\rm Det}{\sf A}}{3} +
\bigl\langle\raisebox{-1pt}{${\sf A}^2$}\bigr\rangle
\frac{2 \langle{\sf AB}\rangle-5\langle{\sf A}\rangle \langle{\sf B}\rangle}{6} -
\frac{\langle{\sf A}\rangle^3 \langle{\sf B}\rangle}{6} ~,
\\ \nonumber \\
\zeta_6^{} &\,=\,& \bigl\langle\raisebox{-1pt}{${\sf A}^2$}\bigr\rangle
\frac{\bigl\langle\raisebox{-1pt}{${\sf B}^2$}\bigr\rangle+\langle{\sf B}\rangle^2}{6} -
\langle{\sf A}\rangle^2\, \frac{\bigl\langle\raisebox{-1pt}{${\sf B}^2$}\bigr\rangle +
7\langle{\sf B}\rangle^2}{6} +
\frac{2\langle{\sf A}\rangle \bigl\langle\raisebox{-1pt}{${\sf AB}^2$}\bigr\rangle +
\langle{\sf B}\rangle \bigl\langle\raisebox{-1pt}{${\sf A}^2\sf B$}\bigr\rangle -
2 \bigl\langle\raisebox{-1pt}{${\sf A}^2{\sf B}^2$}\bigr\rangle}{3} ~,
\\ \nonumber \\
\zeta_7^{} &\,=\,& \bigl\langle\raisebox{-1pt}{${\sf A}^2$}\bigr\rangle
\frac{\bigl\langle\raisebox{-1pt}{${\sf B}^2$}\bigr\rangle-\langle{\sf B}\rangle^2}{12} -
\langle{\sf A}\rangle^2\,\frac{5\bigl\langle\raisebox{-1pt}{${\sf B}^2$}\bigr\rangle +
11 \langle{\sf B}\rangle^2}{12} +
\frac{2 \langle{\sf A}\rangle \bigl\langle\raisebox{-1pt}{${\sf AB}^2$}\bigr\rangle +
\langle{\sf B}\rangle \bigl\langle\raisebox{-1pt}{${\sf A}^2\sf B$}\bigr\rangle -
\bigl\langle\raisebox{-1pt}{${\sf A}^2{\sf B}^2$}\bigr\rangle}{3} ~,
\\ \nonumber \\
\zeta_8^{} &\,=\,& \langle{\sf A}\rangle
\frac{\raisebox{-1pt}{$5\langle{\sf B}\rangle^2$}
+ 3\bigl\langle\raisebox{-1pt}{$\sf B^2$}\bigr\rangle}{6}
- \frac{2\bigl\langle\raisebox{-1pt}{$\sf AB^2$}\bigr\rangle}{3} ~, \hspace{33pt}
\zeta_9^{} \,\,=\,\, \frac{\langle{\sf A}\rangle\bigl\langle\raisebox{-1pt}{$\sf B^2$}\bigr\rangle
- \bigl\langle\raisebox{-1pt}{$\sf AB^2$}\bigr\rangle}{3} ~,
\\ \nonumber \\
\zeta_{10}^{} &\,=\,& \langle{\sf B}\rangle \frac{\raisebox{-1pt}{$5\langle{\sf A}\rangle^2$}
+ \bigl\langle\raisebox{-1pt}{$\sf A^2$}\bigr\rangle}{6}
- \frac{\bigl\langle\raisebox{-1pt}{$\sf A^2B$}\bigr\rangle}{3} ~, \hspace{41pt}
\zeta_{11}^{} \,\,=\,\,
\frac{\bigl\langle\raisebox{-1pt}{$\sf A^2B$}\bigr\rangle
- \bigl\langle\raisebox{-1pt}{$\sf A^2$}\bigr\rangle \langle{\sf B}\rangle}{3} ~,
\\ \nonumber \\
\zeta_{12}^{} &\,=\,& \frac{-\langle{\sf B}\rangle^2}{2}
- \frac{\bigl\langle\raisebox{-1pt}{$\sf B^2$}\bigr\rangle}{6} ~, \hspace{101pt}
\zeta_{13}^{} \,\,=\,\, \frac{4\langle{\sf A}\rangle\langle{\sf B}\rangle
+ \langle{\sf AB}\rangle}{3} ~,
\\ \nonumber \\
\zeta_{14}^{} &\,=\,& \langle{\sf A}\rangle\langle{\sf B}\rangle
- \frac{\langle{\sf AB}\rangle}{3} ~, \hspace{101pt}
\zeta_{15}^{} \,\,=\,\, \frac{-\langle{\sf A}\rangle^2}{2}
- \frac{\bigl\langle{\sf A}^2\bigr\rangle}{6} ~,
\\ \nonumber \\
\zeta_{16}^{} &\,=\,& \frac{2\langle{\sf B}\rangle}{3} ~, \hspace{152pt}
\zeta_{17}^{} \,\,=\,\, \frac{2\langle{\sf A}\rangle}{3} ~.
\end{eqnarray}
The Hermiticity of $\sf A$ and $\sf B$ implies that all the traces and
determinants in $\zeta_{1,2,\cdots,17}^{}$ are purely real, except
$\bigl\langle\raisebox{-1pt}{${\sf A}^2{\sf BAB}^2$}\bigr\rangle$ in $\zeta_1^{}$ which
has an imaginary component
\begin{eqnarray}
J_\xi \,\,=\,\, {\rm Im}\bigl\langle\raisebox{-1pt}{${\sf A}^2{\sf BAB}^2$}\bigr\rangle
\,\,=\,\, \frac{i}{2}\,{\rm Det}[{\sf A,B}]
\end{eqnarray}
obtainable from the Cayley-Hamilton identity
\begin{eqnarray}
[{\sf A},{\sf B}]^3 \,\,=\,\, \openone_{\,}{\rm Det}[{\sf A},{\sf B}] \,+\,
\mbox{$\frac{1}{2}$}\, [{\sf A},{\sf B}] \bigl( \bigl\langle[{\sf A},{\sf B}]^2\bigr\rangle
- \langle[{\sf A},{\sf B}]\rangle^2 \bigr) \,+\,
[{\sf A},{\sf B}]^2\,\langle[{\sf A},{\sf B}]\rangle ~.
\end{eqnarray}
Clearly the reduction of ${\sf A}^2{\sf BAB}^2$ into a sum of matrix products
with lower orders causes the coefficient $\zeta_1$ to gain an imaginary component
equal to~$J_\xi$.
It follows that higher-order matrix products containing ${\sf A}^2{\sf BAB}^2$ will lead to
contributions to the coefficients $\xi_r^{}$ with imaginary parts which are always
proportional to~$J_\xi$

\end{document}